\documentclass[useAMS,usenatbib]{mn2e}
\usepackage{graphicx}
\usepackage{txfonts}
\pdfoutput=1
\title[Characterisation of red-giant stars in public \textit{Kepler} data]{Characterisation of red-giant stars in the public \textit{Kepler} data}

\author[S. Hekker et al. 2010]{S. Hekker$^{1,2}$\thanks{E-mail: S.Hekker@uva.nl; A table with results is available upon request from the first author.}, R.L. Gilliland$^{3}$, Y. Elsworth$^{2}$, W.J. Chaplin$^{2}$, J. De Ridder$^{4}$, D. Stello$^{5}$, 
 \newauthor T. Kallinger$^{6,7}$, K.A. Ibrahim$^{8}$, T.C. Klaus$^{8}$, J. Li$^{9}$\\
$^{1}$Astronomical Institute ``Anton Pannekoek'', University of Amsterdam, PO Box 94249, 1090 GE Amsterdam, The Netherlands\\
$^{2}$School of Physics and Astronomy, University of Birmingham, Edgbaston, Birmingham B15 2TT, UK\\
$^{3}$Space Telescope Science Institute, 3700 San Martin Drive, Baltimore, MD 21218, USA\\
$^{4}$Instituut voor Sterrenkunde, K.U. Leuven, Celestijnenlaan 200D, 3001 Leuven, Belgium\\
$^{5}$Sydney Institute for Astronomy (SIfA), School of Physics, University of Sydney, NSW 2006, Australia\\
$^{6}$Department of Physics and Astronomy, University of British Colombia, 6224 Agricultural Road, Vancouver, BC V6T 1Z1, Canada\\
$^{7}$Institute for Astronomy, University of Vienna, T\"urkenschanzstrasse 17, A-1180 Vienna, Austria\\
$^{8}$Orbital Sciences Corporation/NASA Ames Research Center, Moffett Field, CA 94035, USA\\
$^{9}$SETI Institute/NASA Ames Research Center, Moffett Field, CA 94035, USA\\
  }
  
\begin{document}
\newcommand{\meandnu} {\langle\Delta\nu\rangle}

\date{Received; accepted}
  
\pagerange{\pageref{firstpage}--\pageref{lastpage}} \pubyear{2009}

\maketitle

\label{firstpage}

 
\begin{abstract}
The first public release of long-cadence stellar photometric data collected by the NASA \textit{Kepler} mission has now been made available. In this paper we characterise the red-giant (G-K) stars in this large sample in terms of their solar-like oscillations. We use published methods and well-known scaling relations in the analysis. Just over 70\% of the red giants in the sample show detectable solar-like oscillations, and from these oscillations we are able to estimate the fundamental properties of the stars. This asteroseismic analysis reveals different populations: low-luminosity H-shell burning red-giant branch stars, cool high-luminosity red giants on the red-giant branch and He-core burning clump and secondary-clump giants. 
\end{abstract}


%

\section{Introduction}
The NASA \textit{Kepler} satellite \citep{borucki2009} was launched successfully in March 2009 into an earth-trailing orbit. The satellite is observing $\sim$150\,000 stars simultaneously with high-precision photometry at a near-regular cadence of 29.4 minutes. Data taken during the first 10 day run (Q0), which lasted from 2 May 2009 until 11 May 2009 and the consecutive 33 day run (Q1, 13 May 2009 - 15 June 2009) have been made public\footnote{http://archive.stsci.edu/pub/kepler/lightcurves/tarfiles/}. These data have been processed using the \textit{Kepler} science processing pipeline \citet{jenkins2010a} and have been characterised in terms of their stellar parameters and variability by \citet{ciardi2010}.  
In this work we characterise the red-giant stars using asteroseismology, i.e., the data have been analysed to search for solar-like oscillations. For a recent review of solar-like oscillations in red-giant stars we refer to \citet{hekker2010AN}, while the following papers present recent results from the \textit{Kepler}  \citep[e.g.,][]{bedding2010,huber2010,kallinger2010kepler,hekker2010bin} and CoRoT \citep[e.g.,][]{deridder2009,hekker2009,miglio2009,carrier2010,kallinger2010corot,miglio2010,mosser2010,hekker2010pb} missions. Global oscillation parameters, such as $\nu_{\rm max}$ (the frequency of maximum oscillation power) and $\meandnu$ (the average frequency separation between consecutive overtones of oscillation modes) together with the stellar effective temperature provide valuable information about the stars such as the mass and the radius. These parameters together with other stellar parameters available in the \textit{Kepler} Input Catalogue \citep[KIC,][]{brown2011} give a more complete characterisation of the sample of stars at hand.

\begin{figure}
\begin{minipage}{\linewidth}
\centering
\includegraphics[width=\linewidth]{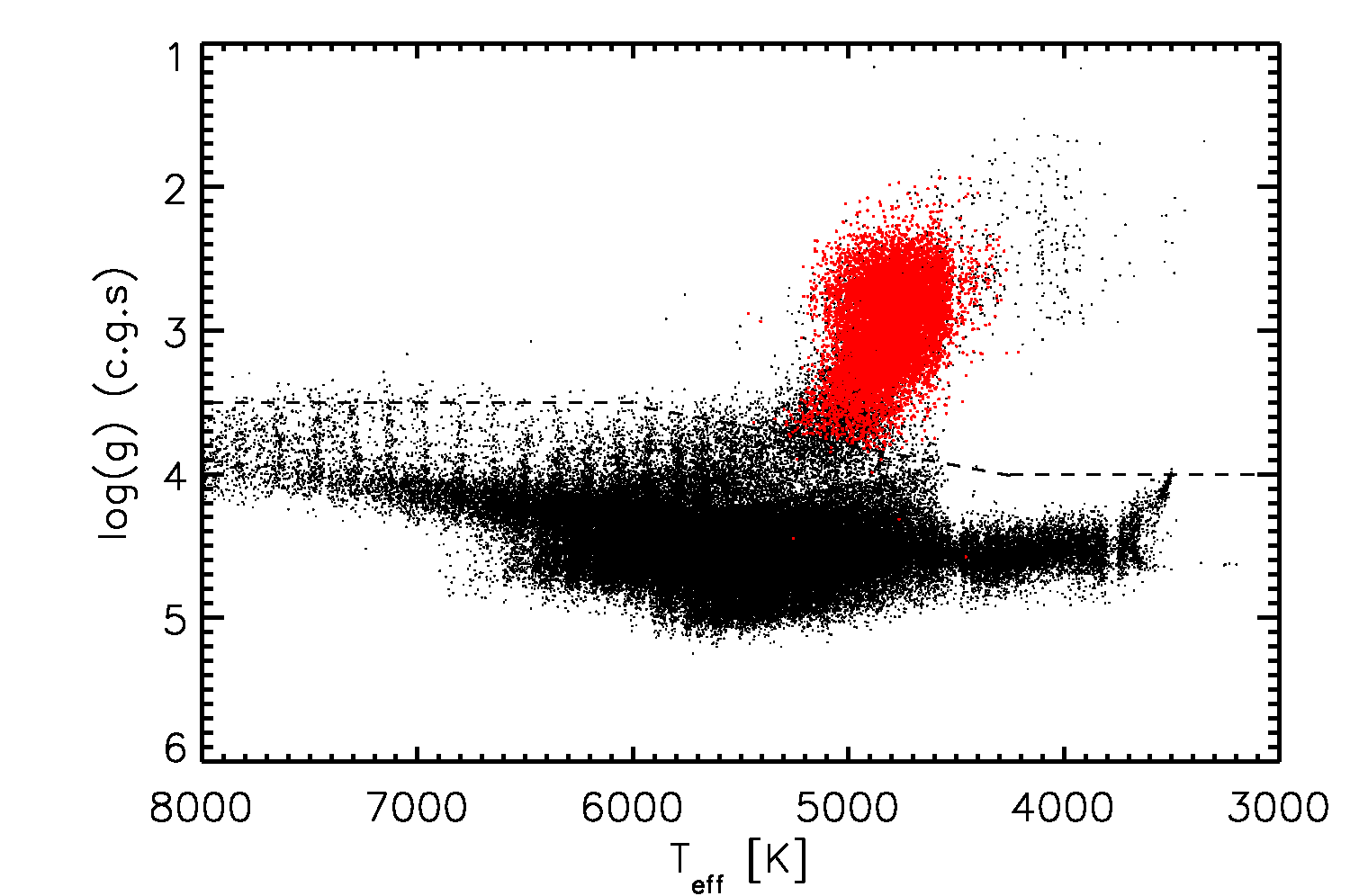}
\end{minipage}
\caption{$\log g$ versus $T_{\rm eff}$ of all $\sim$150\,000 stars for which Q1 public data are available (black) and red giants in which oscillations have been detected (red). The dashed line indicates the division between red giants (low $\log g$) and other stars (high $\log g$) as described by Eq.~\ref{ciardi}.}
\label{loggteff}
\end{figure}

\section{Target selection}
The Q0 timeseries alone are considered to be too short to determine global oscillation parameters, and only those stars for which Q1 timeseries are available have been analysed. This comprised a sample of 150\,597 stars in total. From this sample of stars a subset of 16\,511 red-giant stars have been selected using the definition of red giants provided by \citet{ciardi2010}:
\begin{equation}
\log g \leq 5.2-2.8\times10^{-4} T_{\rm eff}/\rm K,
\label{ciardi}
\end{equation}
for the range 4250~K~$<T_{\rm eff} < $~6000~K, and $\log g \leq 4$ for stars with $T_{\rm eff} < $~4250~K (see Fig.~\ref{loggteff}). 

The sample of stars investigated here is distinct from the set of oscillating red giants in \textit{Kepler} data discussed in the several papers mentioned in the introduction. For the earlier papers, targets selected either as astrometric, red-giant control stars (about 1\,000 uniformly distributed over the field of view with an apparent magnitude in the \textit{Kepler} bandpass ($K_p$) ranging from 11.3 to 12.1), or a comparably large number of generally brighter giants that had been selected for observation by the Kepler Asteroseismic Science Consortium (KASC) have been used. In both cases these time series for Q1 were not included in the public data set analysed in this paper. These public data come from the \textit{Kepler} science team targets selected on the basis of allowing planets to be detected if transiting these stars \citep{brown2011}. 

\section{Results}
\subsection{Oscillation parameters}
Power spectra of the red-giant sample have been analysed using the analysis methods described by \citet{hekker2010pipe}. In short, the computation of $\meandnu$ is based on the power spectrum of the power spectrum in which the $\meandnu$/2 and $\meandnu$/4 peaks are determined. The centroids of these peaks have been computed as the power weighted centroid and the uncertainties have been computed as the standard deviation of grouped data, taking into account the oversampling.  $\meandnu$ is computed as the weighted average of the two measured features. $\nu_{\rm max}$ is determined as the centre of a Gaussian fit through the oscillation power excess in the binned power spectrum. The uncertainty is estimated as the standard deviation of this parameter resulting from the Gaussian fit.

We already had at our disposal datasets on red-giant KASC targets that are of the same length as the public datasets. These KASC datasets have 
been analysed by several independent teams \citep[][and references therein]{hekker2010comp}, from which, in general, consistent results were 
obtained. Additionally, it was also clear that no significant biases were present in the results from the methods used here. We can therefore have confidence that the results presented here, from one method, should be robust. \citet{verner2011} have investigated the uncertainties derived by different methods from artificial data and observations of approximately the same length as the public data discussed here. For each method, multiplicative correction factors for the uncertainties in $\meandnu$ and $\nu_{\rm max}$ have been computed as 2.03 and 1.09 respectively, which we have applied to the uncertainties derived with the automated method.

\begin{figure}
\begin{minipage}{\linewidth}
\centering
\includegraphics[width=\linewidth]{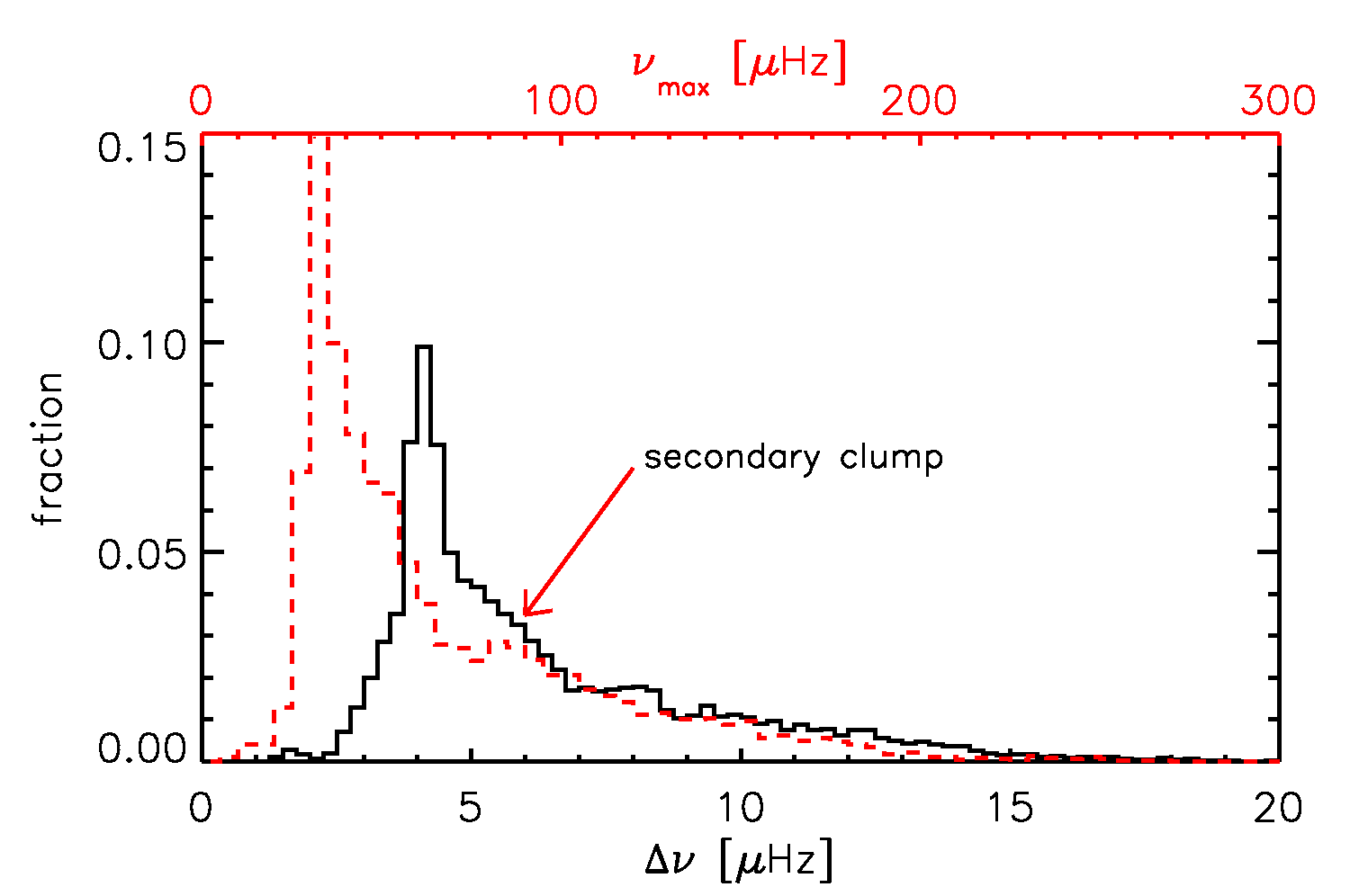}
\end{minipage}
\hfill
\begin{minipage}{\linewidth}
\centering
\includegraphics[width=\linewidth]{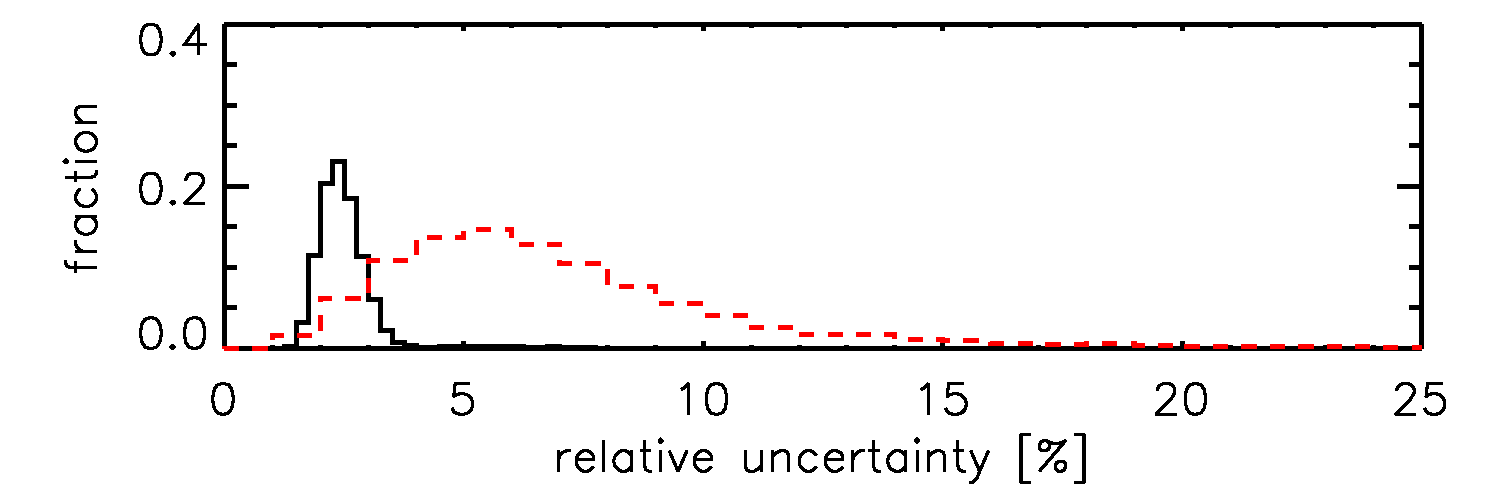}
\end{minipage}
\caption{Top: distribution of $\nu_{\rm max}$ (red dashed line and top-axis)  and $\meandnu$ (black solid line and bottom axis). Bottom: distribution of relative corrected uncertainties in percent of $\nu_{\rm max}$ (red dashed line) and $\meandnu$ (black).}
\label{dnunumaxhisto}
\end{figure}

\citet{hekker2010comp} investigated the influence of the realization noise, i.e., a measure of the variation in the observed parameters due to the stochastic nature of the oscillations, on the derived values of $\meandnu$ and $\nu_{\rm max}$. They analysed different realizations of simulated data with the same input parameters using different methods. The simulated time series have a similar timespan as the data investigated here. For these timeseries, \citet{hekker2010comp} find that scatter due to realization noise is non-negligible and can be at least as important as the internal uncertainty of the result due to the method used. For the methods used in this work, the scatter in the results of the different realizations was in general smaller than the quoted uncertainties. Therefore, we are confident that the results we obtain here for one realization are precise within the quoted uncertainties.

In Fig.~\ref{dnunumaxhisto}, the fractional distributions of $\nu_{\rm max}$ and $\meandnu$ values are shown as well as the distributions of the corrected uncertainties. The large number of stars with 30~$\mu$Hz~$<\nu_{\rm max}<$~50~$\mu$Hz and $\meandnu$ around 4~$\mu$Hz are He-burning red-clump stars \citep{miglio2009}. The additional smaller secondary hump at $\nu_{\rm max}$ between roughly 60 and 110 $\mu$Hz, indicated with the red arrow, can be attributed to the secondary-clump \citep{miglio2009}. The secondary-clump consists of He-burning stars with masses high enough to have ignited He in a non-degenerate core.\newline

\begin{figure}
\begin{minipage}{\linewidth}
\centering
\includegraphics[width=\linewidth]{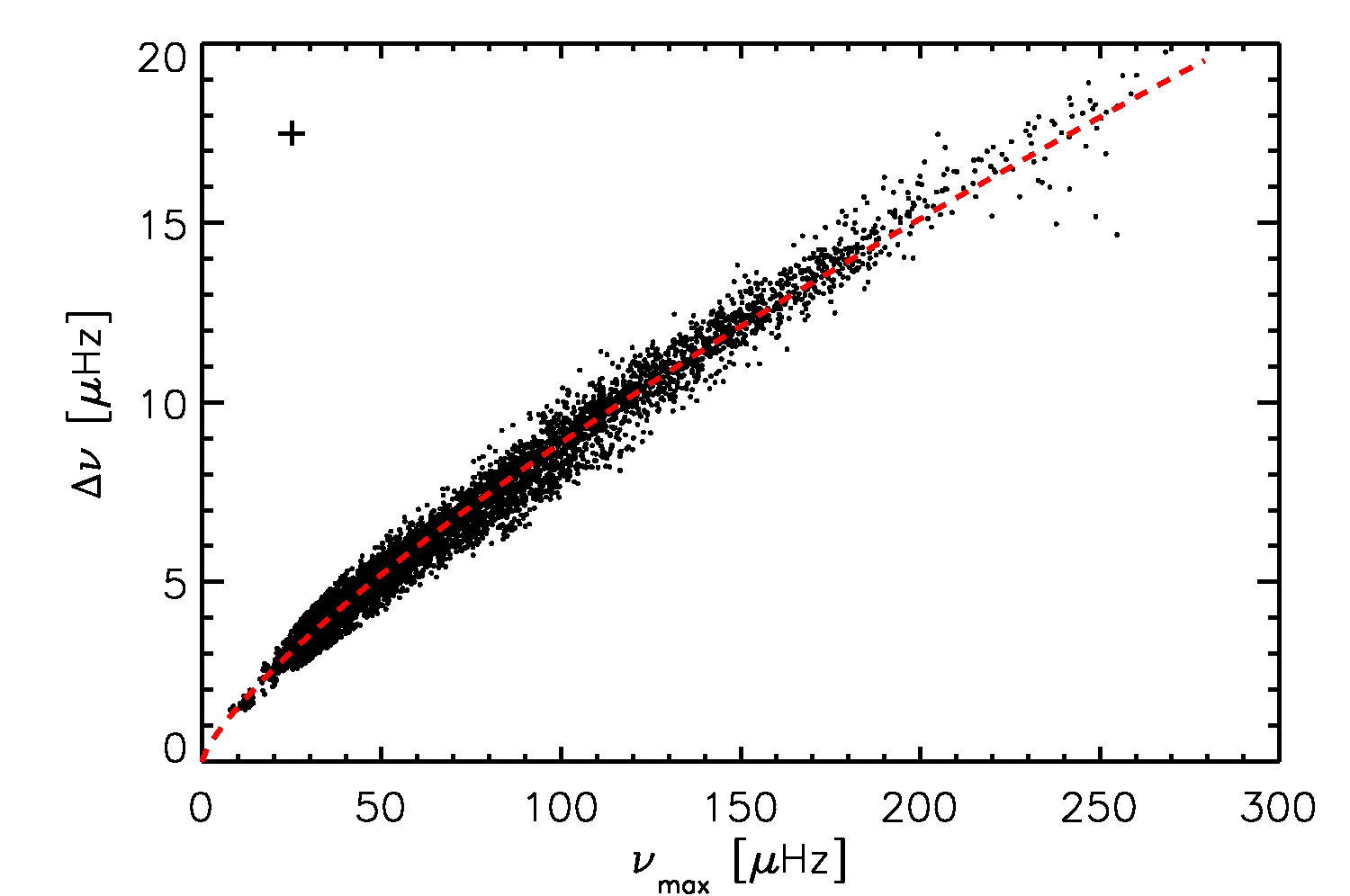}
\end{minipage}
\caption{$\meandnu$ versus $\nu_{\rm max}$ for the oscillating red giants with the characteristic uncertainty shown with a cross in the left top corner. The red dashed line indicates the power-law fit through the results (see text).}
\label{dnunumax}
\end{figure}

We expect a correlation between $\meandnu$ and $\nu_{\rm max}$ \citep{hekker2009,stello2009,mosser2010}. This correlation is shown in Fig.~\ref{dnunumax} together with a polynomial fit of the form:
\begin{equation}
\meandnu = a  \nu_{\rm max}^b,
\label{numaxdnufit}
\end{equation}
with [$a$,$b$] = [0.257$\pm$0.002, 0.769$\pm$0.002], in which the uncertainties indicate one sigma values of the fit. These values are consistent with earlier results presented by \citet{huber2010} for a sample including both red giants and main-sequence stars. For red giants, the value of $a$ from different methods as mentioned by \citet{huber2010} are in general slightly higher than the value for $a$ we found here, while the values of $b$ in \citet{huber2010} are slightly lower than the value quoted here. We investigated the difference in the resulting fits and found that the relative difference in the prediction of $\meandnu$ from $\nu_{\rm max}$ from different fits ranges between 0.3 and 1.5\% for stars with $\nu_{\rm max}$~$>$~10~$\mu$Hz. This is smaller than the uncertainties with which the values can generally be derived. We note that a spread in the correlation in Eq.~\ref{numaxdnufit} is expected mainly due to stellar masses  \citep[see e.g.][]{hekker2010wg2,huber2010,stello2009}. In this study we investigate a larger sample spanning a larger parameter space (see also Fig.~\ref{fracloggteffkasc}) compared to earlier studies.

\subsection{Stellar parameters}
\begin{figure*}
\begin{minipage}{0.45\linewidth}
\centering
\includegraphics[width=\linewidth]{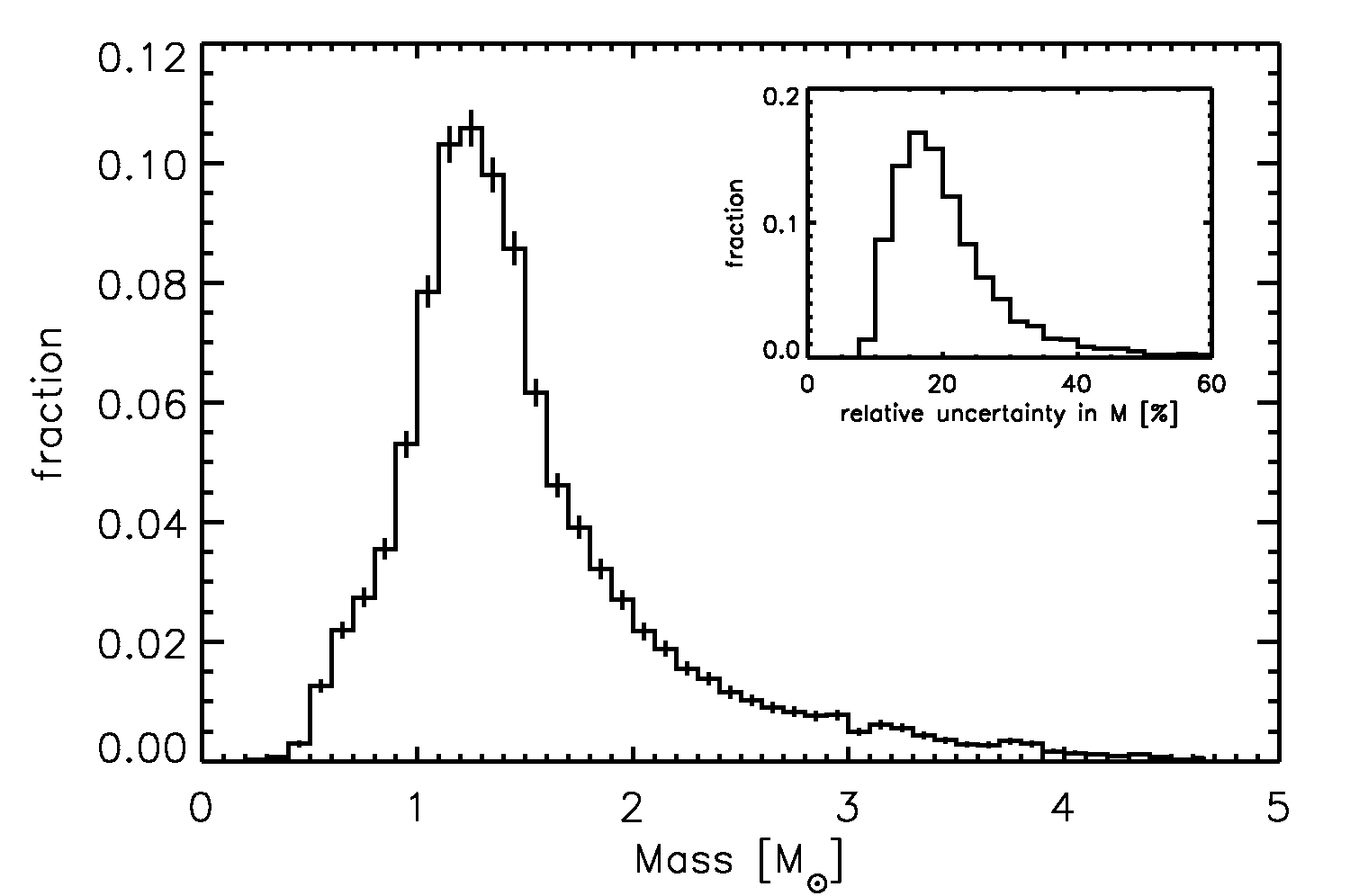}
\end{minipage}
\begin{minipage}{0.45\linewidth}
\centering
\includegraphics[width=\linewidth]{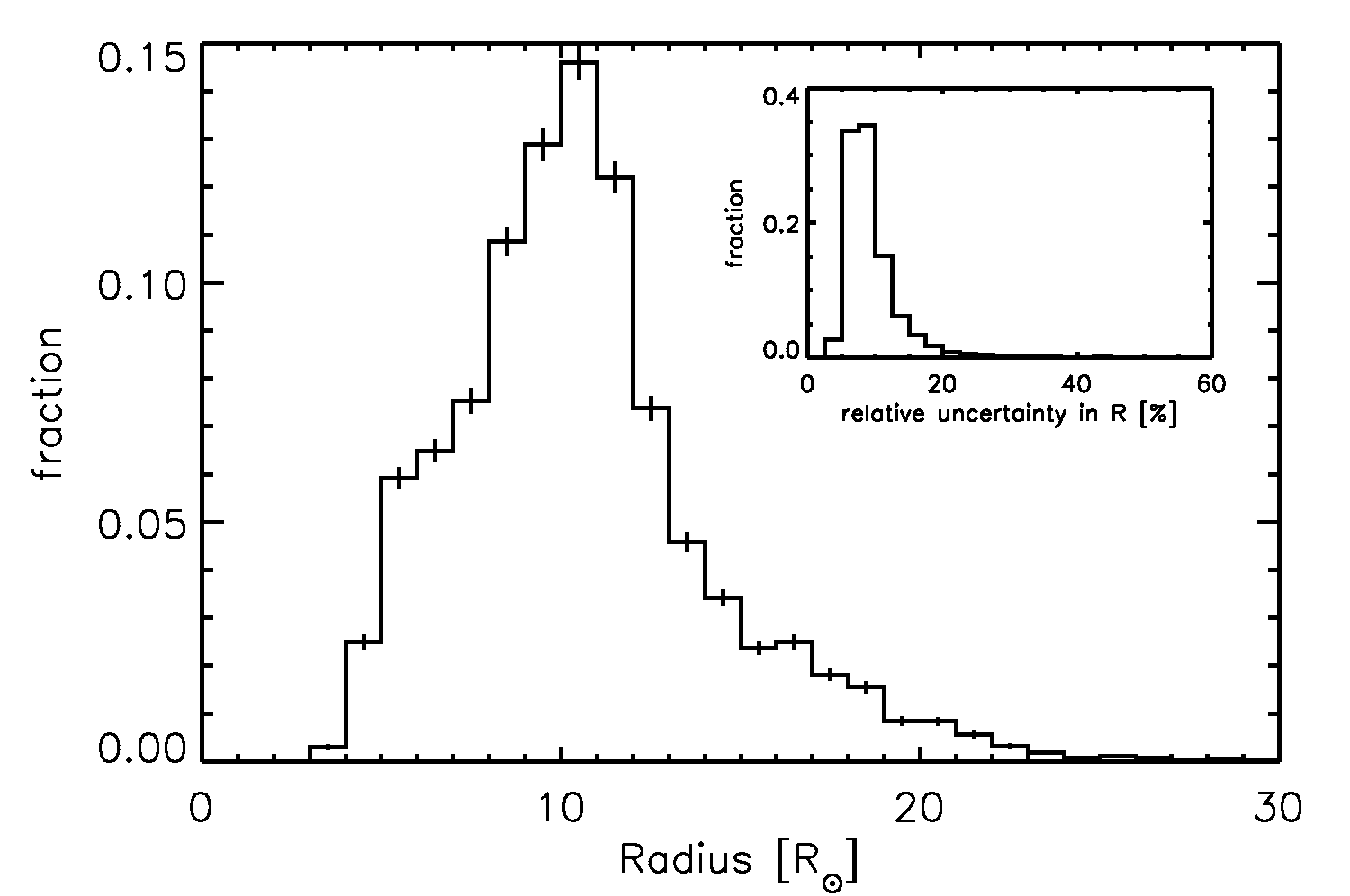}
\end{minipage}
\begin{minipage}{0.45\linewidth}
\centering
\includegraphics[width=\linewidth]{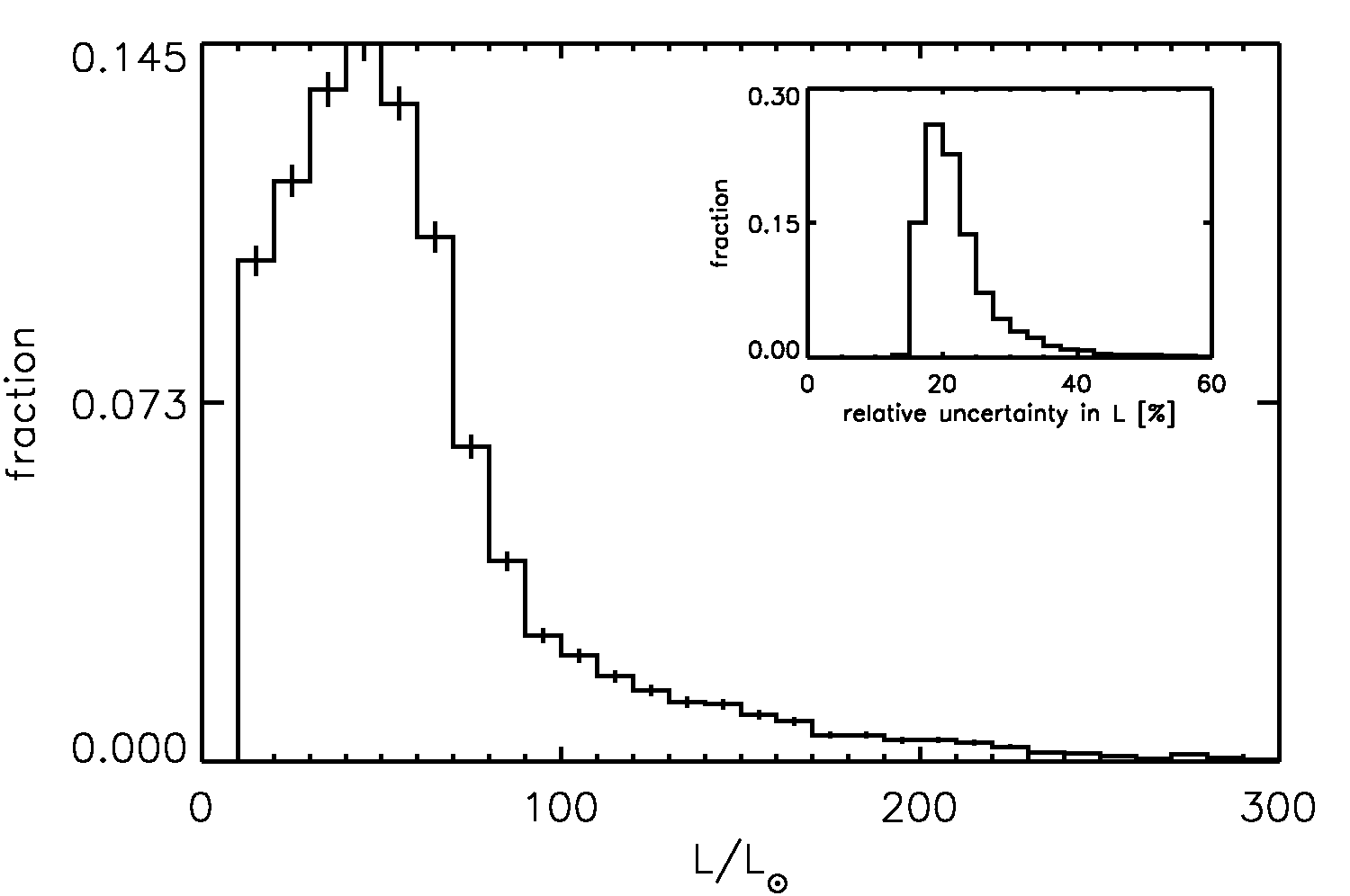}
\end{minipage}
\begin{minipage}{0.45\linewidth}
\centering
\includegraphics[width=\linewidth]{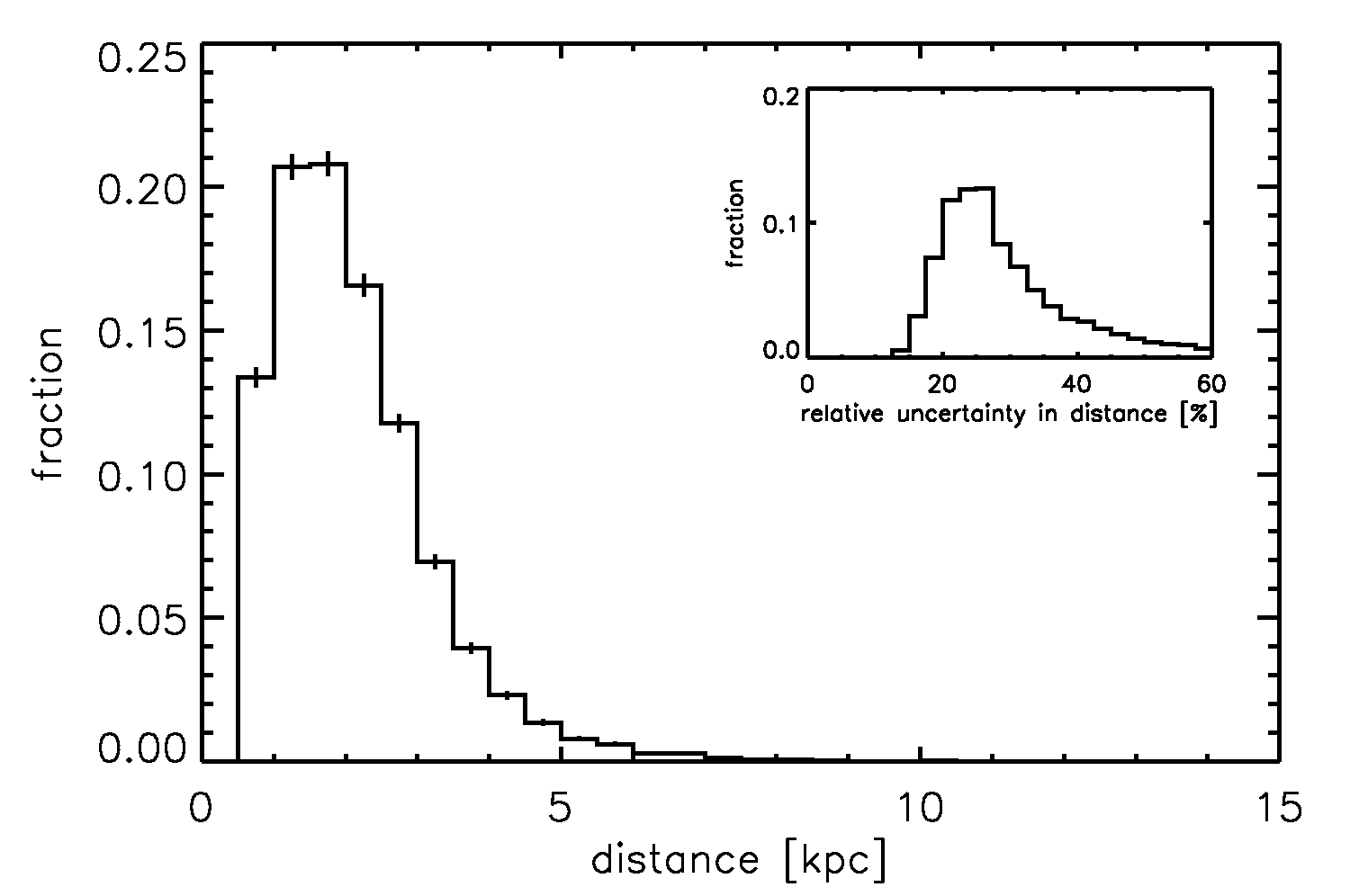}
\end{minipage}
\caption{Fractional distributions of asteroseismic masses (left top), radii (right top), luminosities (left bottom) and distances (right bottom) of the red giants with detected solar-like oscillations in the public data. Uncertainties in the histograms are computed assuming Poisson statistics. The distribution  of the relative uncertainties of the parameters are shown in the insets in each panel.}
\label{MRL}
\end{figure*}

Following the approach by \citet{kallinger2010corot} we use the values of $\meandnu$, $\nu_{\rm max}$ and the effective temperatures from the \textit{Kepler} Input Catalog in combination with the scaling relations as described by \citet{kjeldsen1995} to compute the masses and radii of the stars directly:
\begin{equation}
\nu_{\rm max}=\nu_{\rm max \odot}\frac{M/M_{\odot}}{(R/R_{\odot})^2 \sqrt{T_{\rm eff}/T_{\rm eff \odot}}} \mu {\rm Hz},
\label{numax}
\end{equation}
\begin{equation}
\Delta\nu = \Delta\nu_{\odot}\sqrt{\frac{M/M_{\odot}}{(R/R_{\odot})^3}}  \mu {\rm Hz},
\label{dnu}
\end{equation}
using the solar values, $\nu_{\rm max \odot}$~=~3120~$\mu$Hz,  $\Delta \nu_{\odot}$~=~134.9~$\mu$Hz and $T_{\rm eff \odot}$~=~5777~K, as determined by \citet{kallinger2010kepler}.
From the radii and $T_{\rm eff}$ we computed the luminosity as $L$~$\propto$~$R^2 T_{\rm eff}^4$. Using this luminosity, the apparent \textit{Kepler}  magnitude and the extinction coefficient ($A_V$, available in the KIC) we computed the distances of the stars. The uncertainties for $R$, $M$ and $L$ are computed from propagation of uncertainties, assuming the parameters are uncorrelated. The uncertainty in $T_{\rm eff}$ from the \textit{Kepler} Input Catalog \citep{brown2011} has been estimated to be 150~K. Typical uncertainties in the \textit{Kepler} magnitude are quoted by \citet{brown2011} to be 0.03 mag and we estimate the uncertainty in the reddening to be 0.1. Histograms of masses, radii, luminosities, distances and their respective relative uncertainties are shown in Fig.~\ref{MRL}.

We note here that we did not apply any correction in the derivation of the masses and radii of the stars as those applied by \citet{mosser2010}. These corrections were not implemented for several reasons: 1) No signifiant biases are expected in $\meandnu$ and $\nu_{\rm max}$. Also, no biases are expected in $T_{\rm eff}$ from KIC parameters, as was shown by \citet{kallinger2010kepler} in a comparison between the asteroseismic and the KIC $T_{\rm eff}$. 2) The $T_{\rm eff}$ for a few \textit{Kepler} stars obtained from high-resolution, high signal-to-noise spectra are consistent with $T_{\rm eff}$ from KIC and asteroseismology \citep{bruntt2011}. 3) The position of the stars in the H-R diagram are in agreement with the theoretical population study by \citet{miglio2009}. Despite the consistency of our results with several other works, we do see a bias in effective temperature and luminosity compared to the CoRoT results \citep[see Fig.~15 of][]{mosser2010}. Because of the consistency of $T_{\rm eff}$ from the KIC and spectroscopy and the agreement with the theoretical population study, the higher \textit{Kepler} effective temperatures are preferred. We note however, that the existence of a bias arising from the scaling relations can not be excluded.

The mass distribution has one clear peak between 1 and 2 M$_{\odot}$, while the distribution of the radii shows indications of a dominant peak located at $\sim$11~R$_{\odot}$. These peaks in the distributions are most likely due to stars in the red clump, i.e. He-burning stars \citep[see][for a population study of red giants]{miglio2009}. The radius distribution shows an additional shoulder between roughly 5 and 8~R$_{\odot}$, which consists most likely of less evolved H-shell burning red giants ascending the giant branch. 

The majority of the observed red giants have luminosities between 10 and 100~L$_{\odot}$ and reside at a distance between 1 and 5 kpc. The steep fall off for low luminosities / shorter distances and the gradual fall off in the number of stars at higher luminosities / larger distances are due to the magnitude range accessible with the \textit{Kepler} satellite. 

Note that the masses, radii, luminosities and distances computed from the direct method have relatively large uncertainties. For the radius, the majority of the stars have a relative uncertainty of less than 10\%, while the distributions of the relative uncertainties in the mass, luminosity and distance peak at about 16\%, nearly 20\% and 25\%, respectively. These uncertainties are expected to broaden the distributions, but have minimal influence on the location of dominant peaks in the distributions \citep[see][for an in-depth investigation]{gai2010}. 

\section{Discussion}
For 71\% (11\,805 out of 16\,511) of the red giants, as defined by \citet{ciardi2010}, solar-like oscillations could be detected. Comparing this fraction with the performance of CoRoT, we see that \citet{mosser2011} detect oscillations in 75\% of the red-giant candidates brighter than 13$^{\rm th}$ mag. These results are broadly consistent with each other. 

What could be the reason that we do not detect oscillations in the other 29\% of the stars? This could be either due to analysis or observation biases, or due to intrinsic effects in the stars. All stars have been analysed with an automated method \citep{hekker2010pipe} and checked by eye. This makes sure that the detection of solar-like oscillations is not significantly influenced by the automation. The observations are such that we are able to observe oscillations with $\nu_{\rm max}$ only in the range between roughly 10 and 250 $\mu$Hz.  The lower limit is due both to the frequency resolution and to the presence of low frequency signatures, among which granulation, present in the power spectra. The upper limit is set by the Nyquist frequency of $\sim$283~$\mu$Hz. From $\log g$ and effective temperature we computed a prediction for $\nu_{\rm max}$ for the stars for which no oscillations were detected. For $\sim$4200 stars the predicted $\nu_{\rm max}$ were indeed outside the frequency range for which the current data are sensitive. Furthermore, artefacts such as the signature of the cycling of the heater on one of the \textit{Kepler} reaction wheels \citep[as already reported by][and in the \textit{Kepler} data release notes]{hekker2010comp} might also hamper the detection of the oscillations. Contamination from other instrumental effects or photon-shot noise may also be relevant. Additionally, we found during the manual check that there are some binaries and classical oscillators present in the sample which features hamper the detection of solar-like oscillations. Note that for binaries the KIC estimate of $T_{\rm eff}$ and $\log g$ are rather doubtful. \citet{prsa2010} give a conservative estimate of only 1.2\% eclipsing binaries in the sample.


\subsection{Populations}
\begin{figure*}
\begin{minipage}{0.45\linewidth}
\centering
\includegraphics[width=\linewidth]{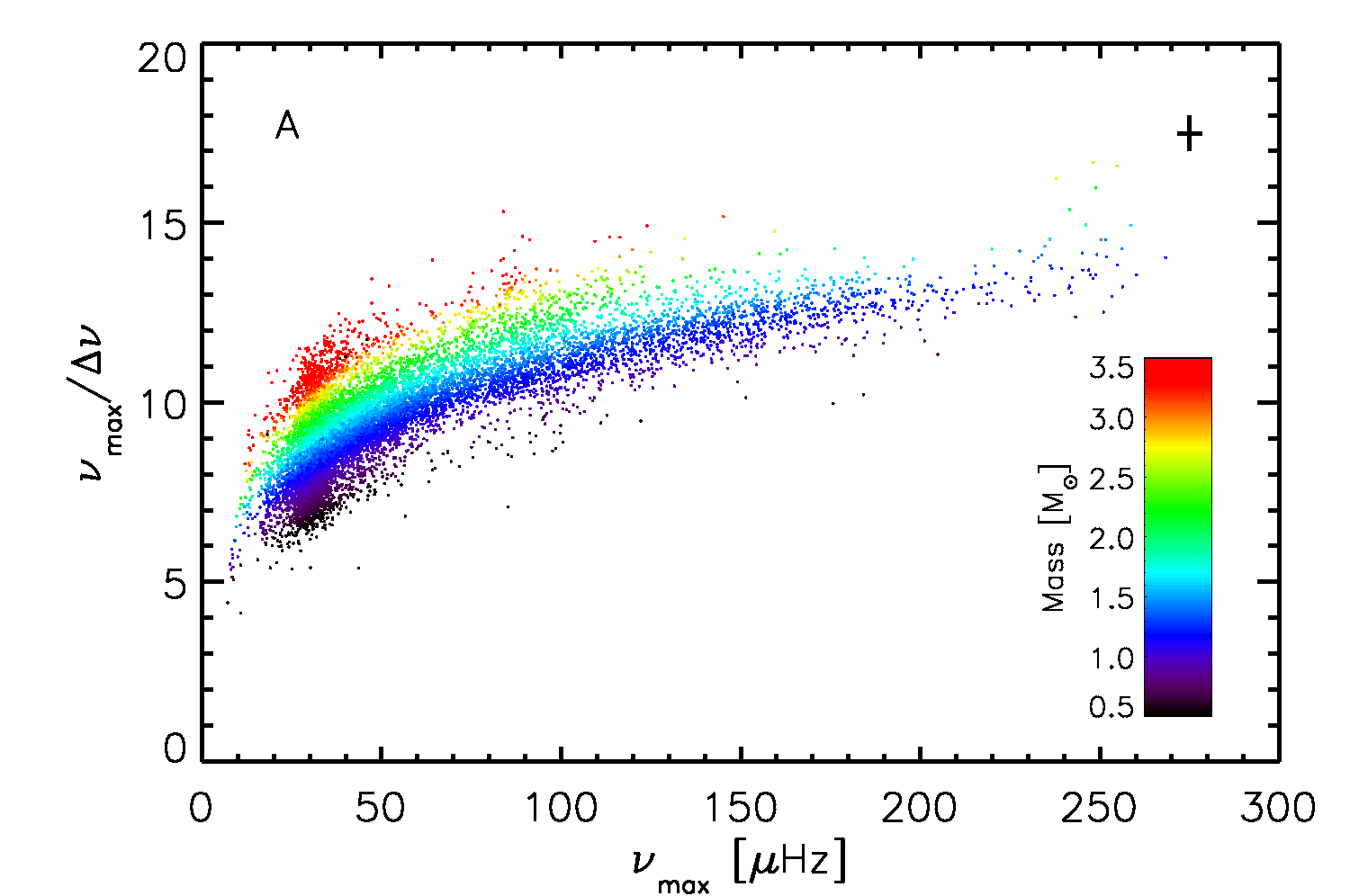}
\end{minipage}
\begin{minipage}{0.45\linewidth}
\centering
\includegraphics[width=\linewidth]{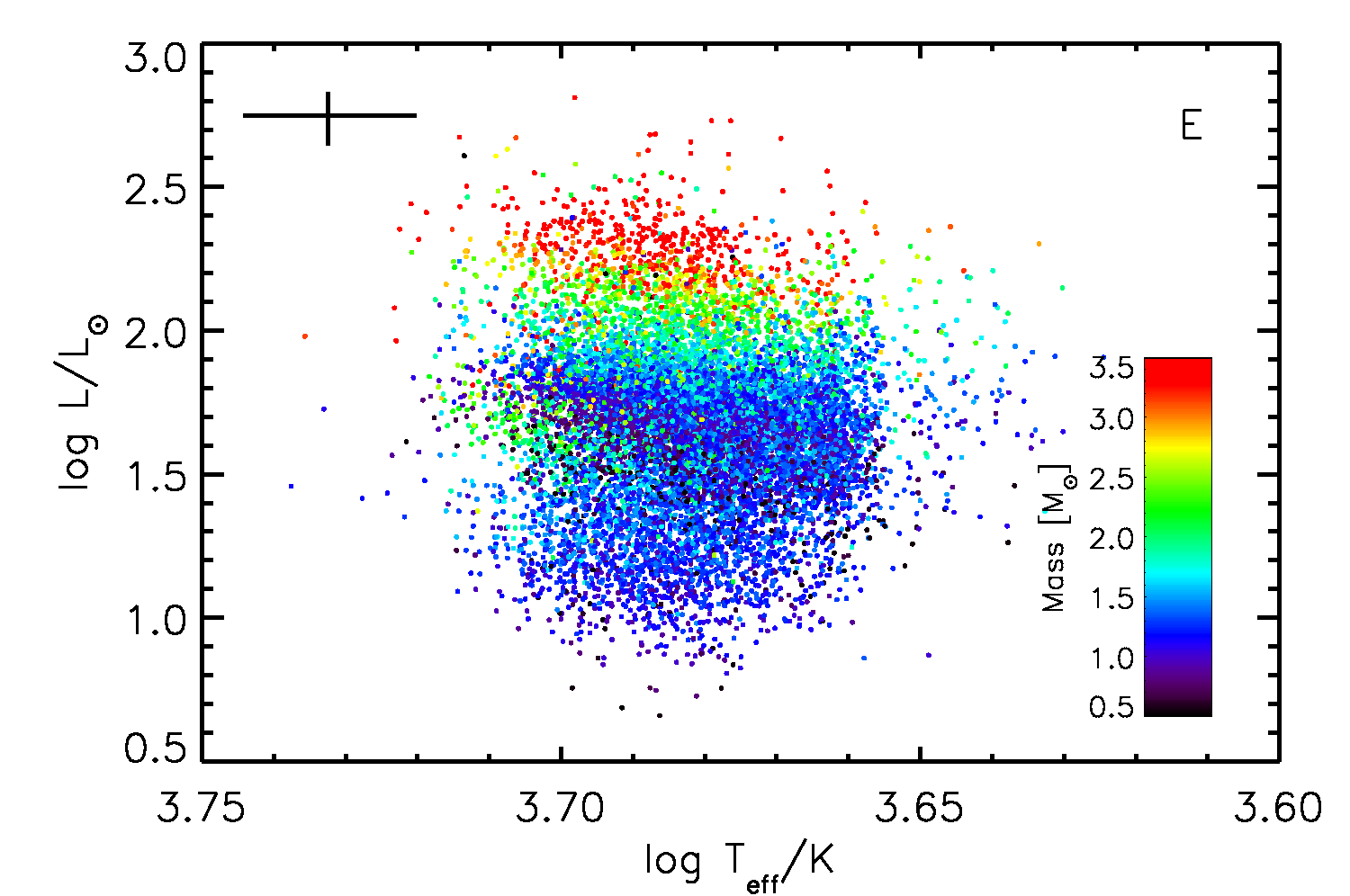}
\end{minipage}
\begin{minipage}{0.45\linewidth}
\centering
\includegraphics[width=\linewidth]{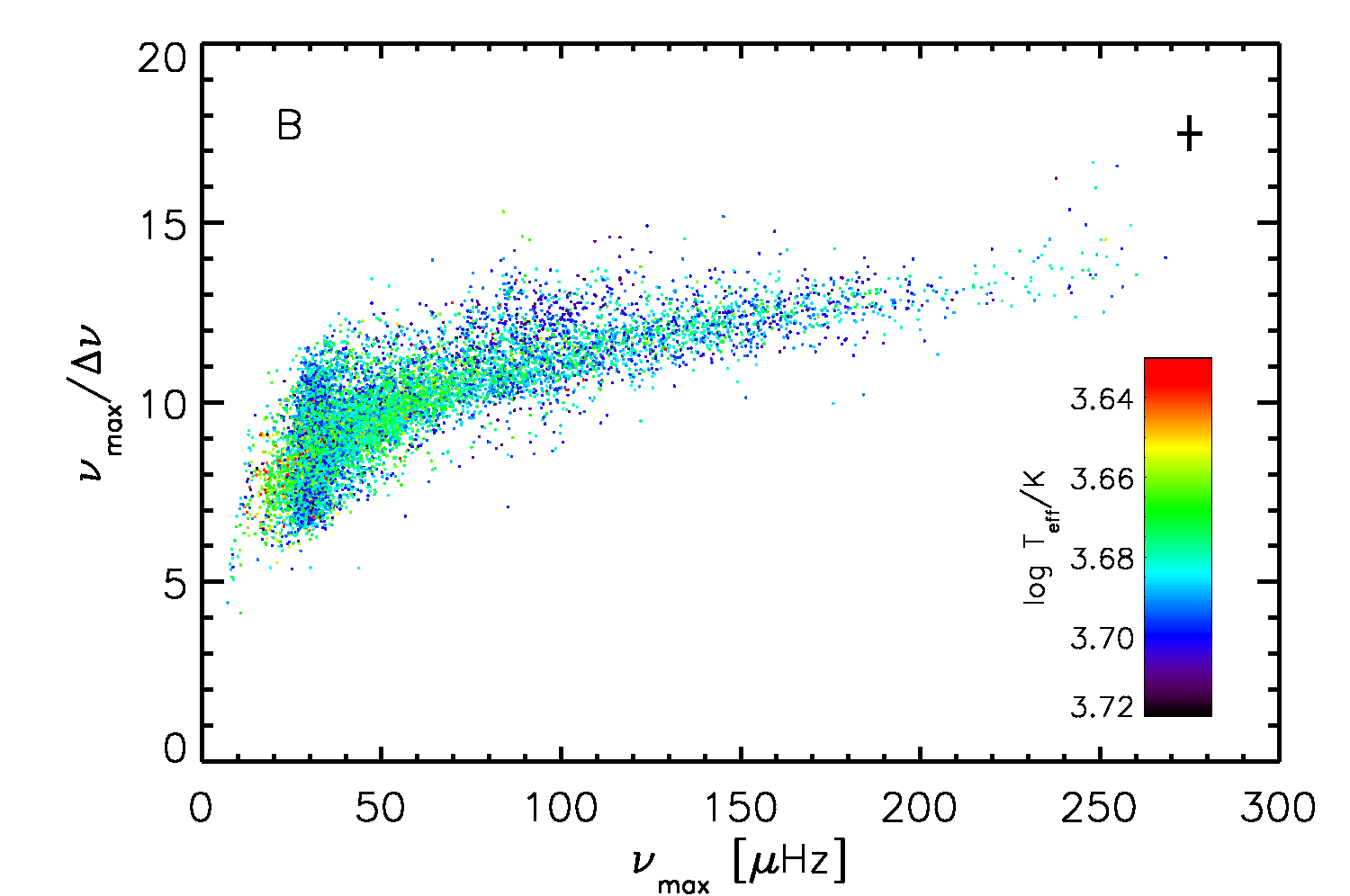}
\end{minipage}
\begin{minipage}{0.45\linewidth}
\centering
\includegraphics[width=\linewidth]{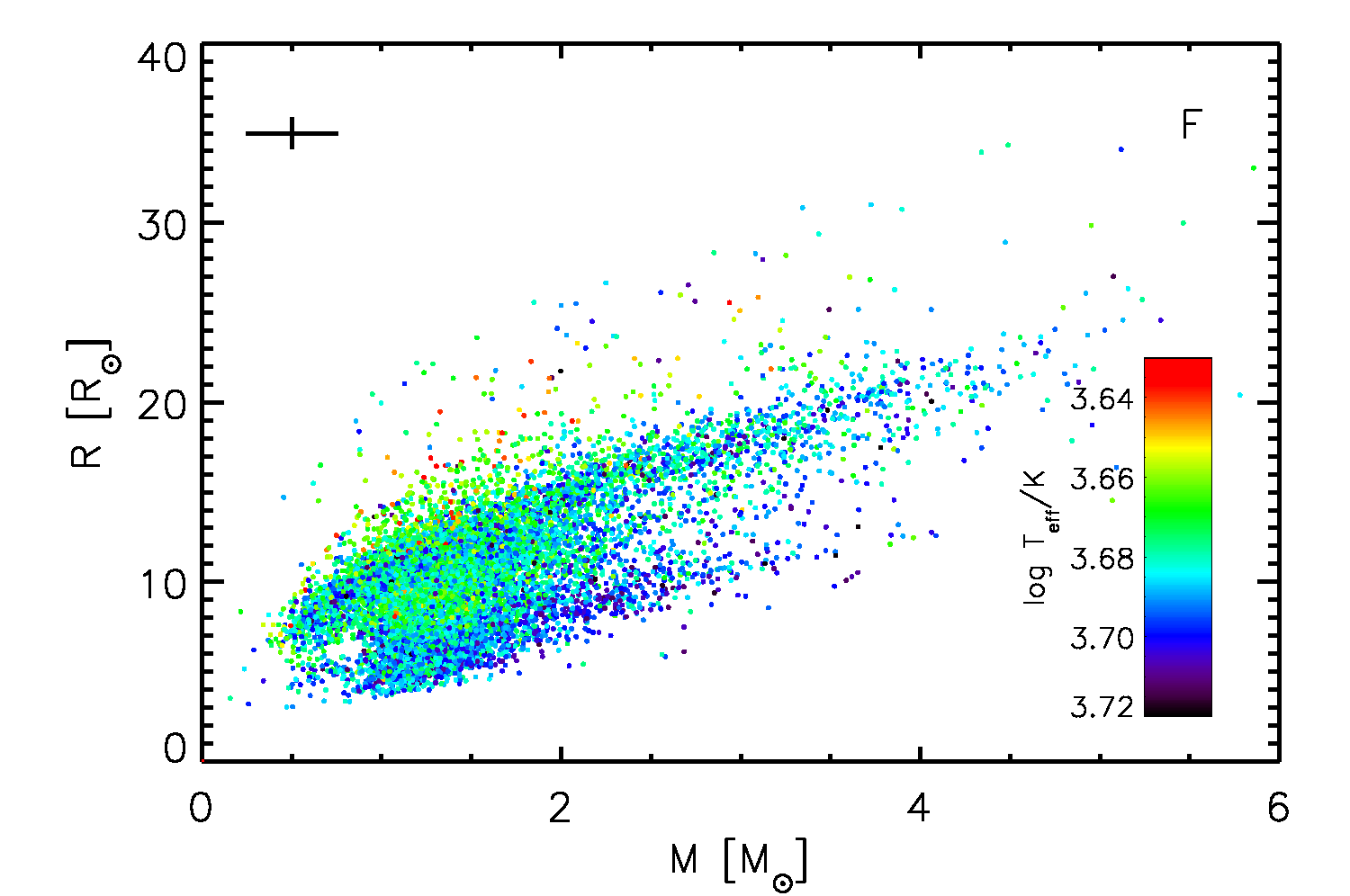}
\end{minipage}
\begin{minipage}{0.45\linewidth}
\centering
\includegraphics[width=\linewidth]{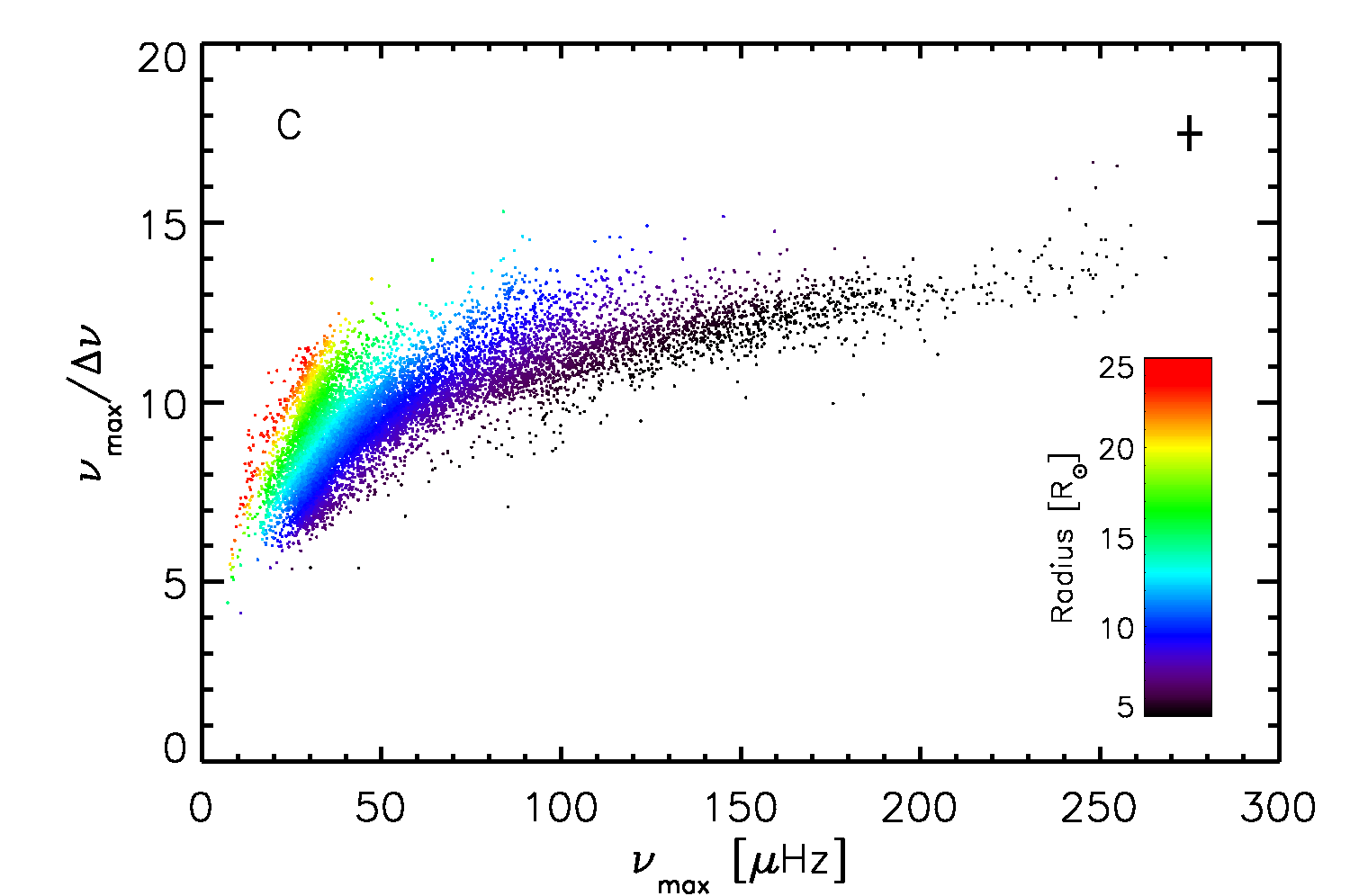}
\end{minipage}
\begin{minipage}{0.45\linewidth}
\centering
\includegraphics[width=\linewidth]{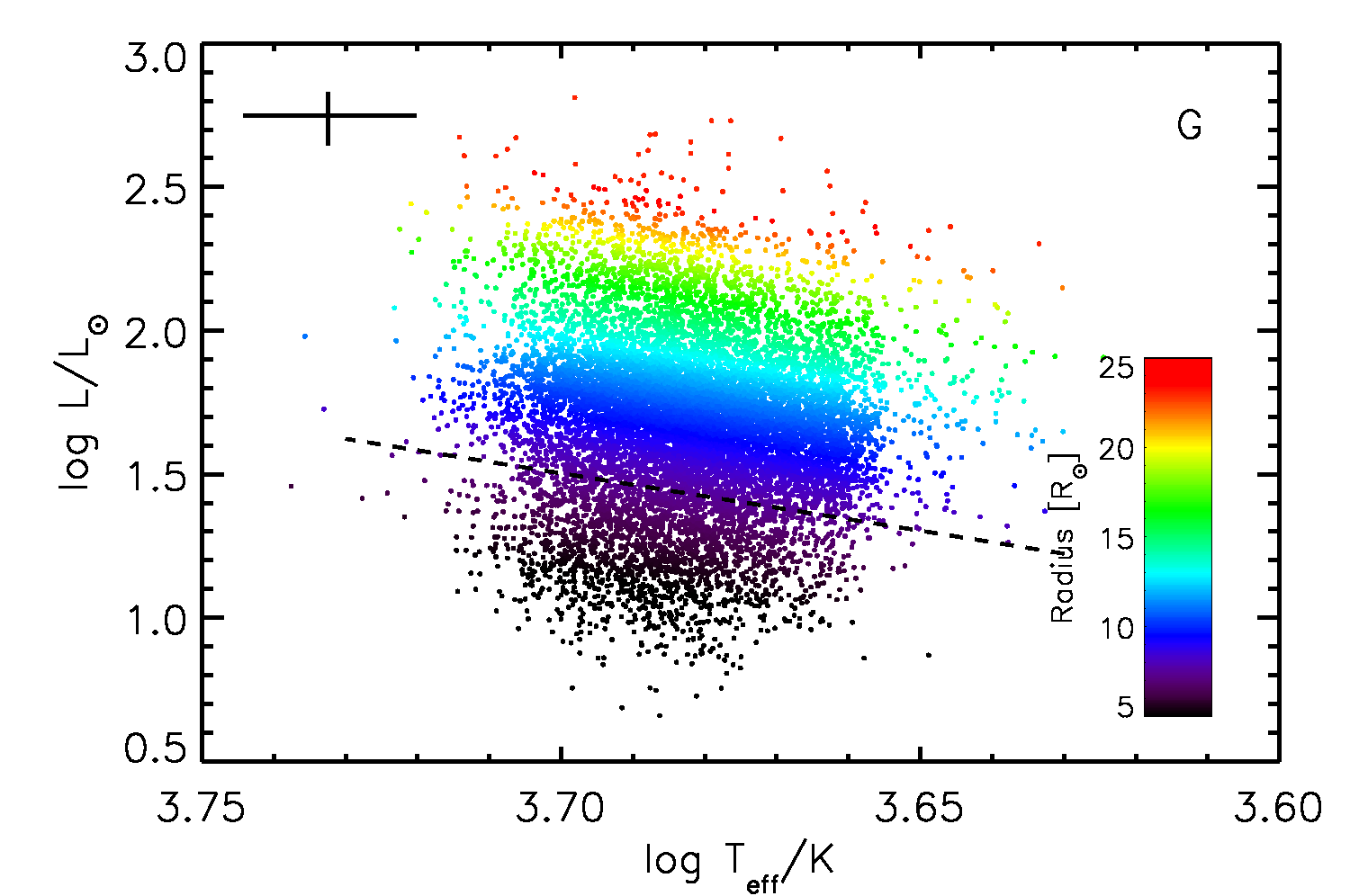}
\end{minipage}
\begin{minipage}{0.45\linewidth}
\centering
\includegraphics[width=\linewidth]{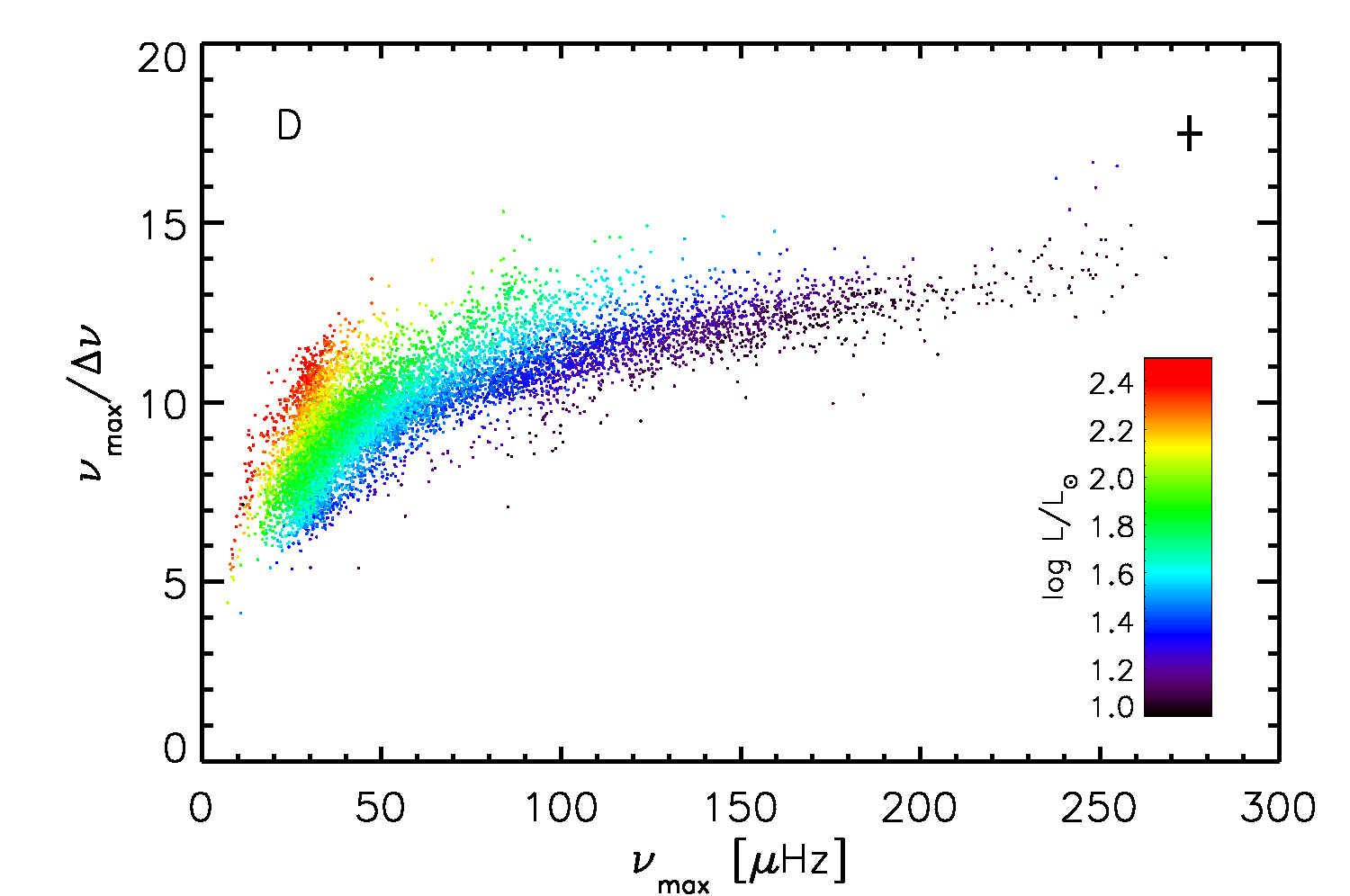}
\end{minipage}
\begin{minipage}{0.45\linewidth}
\centering
\includegraphics[width=\linewidth]{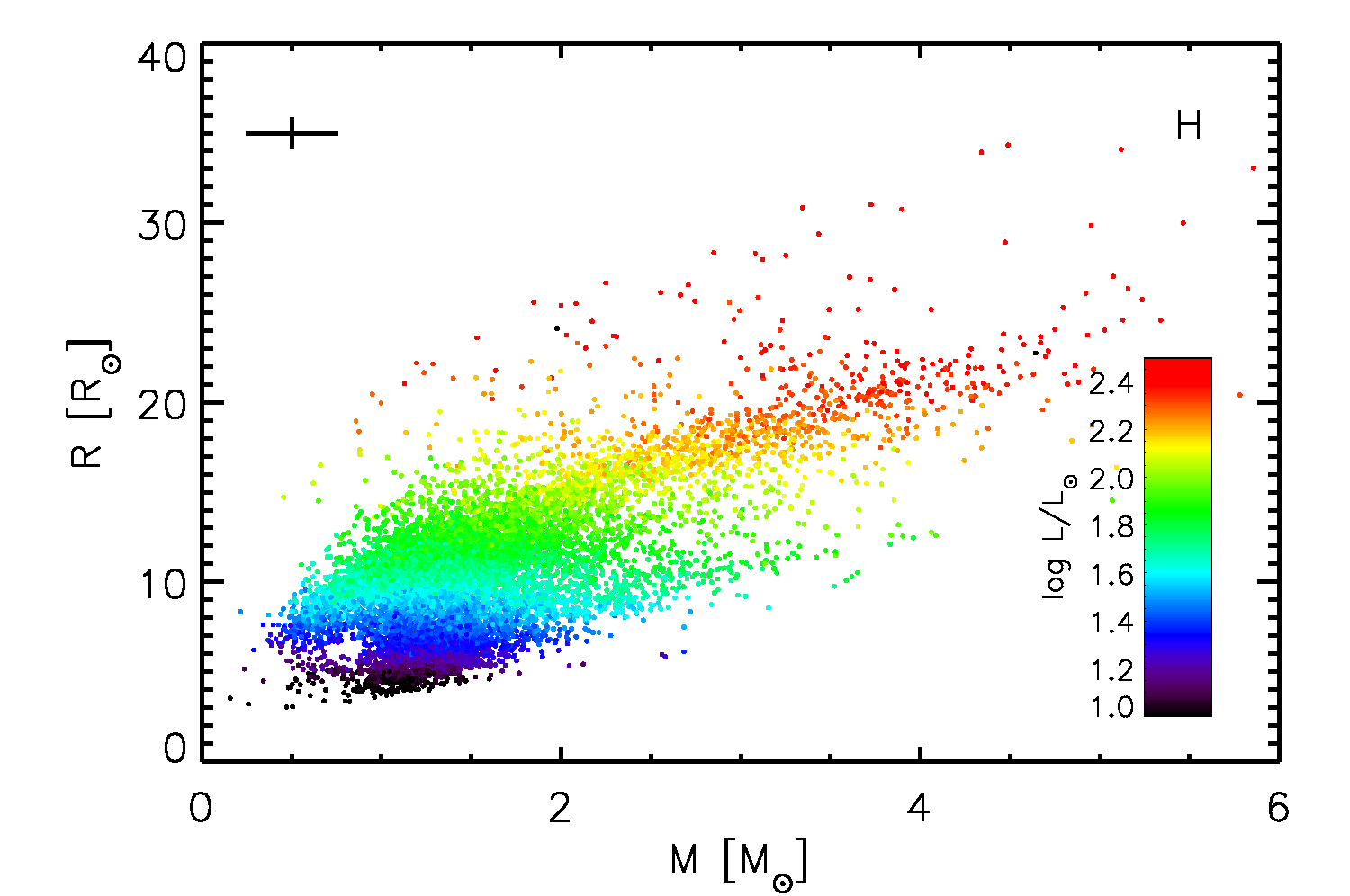}
\end{minipage}
\caption{$\nu_{\rm max}$ versus $\nu_{\rm max}/\meandnu$ diagrams (left) and H-R diagrams and mass versus radius diagrams (right) of the public red giants with detected oscillations. The colour-coding indicates from top to bottom mass, effective temperature, radius and luminosity. The highest and lowest values in the colour scale are lower and upper limits, respectively, to enhance the colour contrast. Characteristic uncertainties are indicated with a cross in the corner of each panel. The black dashed line in panel G indicates a radius of 7.5~R$_{\odot}$ (see text).}
\label{res}
\end{figure*}

Despite the relatively large uncertainties in the stellar parameters (mass, radius, luminosity) of individual stars the statistically significant number of stars allows us to perform a global investigation of the parameters of the complete sample and investigate the population of stars.
For this investigation of the population of red giants for which oscillations have been detected we show $\nu_{\rm max}$ versus $\nu_{\rm max}/\meandnu$ in Fig.~\ref{res}, in the same way as \citet{huber2010}, but now with colour-coding for mass, radius, luminosity and effective temperature also included. The mass, radius and luminosity of the stars in this diagram show smooth transitions with the global oscillation parameters, as predicted from stellar models.  However the trends in global oscillation parameters appear to be almost independent of effective temperature. In the same figure, we also show H-R diagrams colour-coded for radius and mass, and mass versus radius diagrams colour-coded for luminosity and effective temperature. 


\begin{figure*}
\begin{minipage}{0.45\linewidth}
\centering
\includegraphics[width=\linewidth]{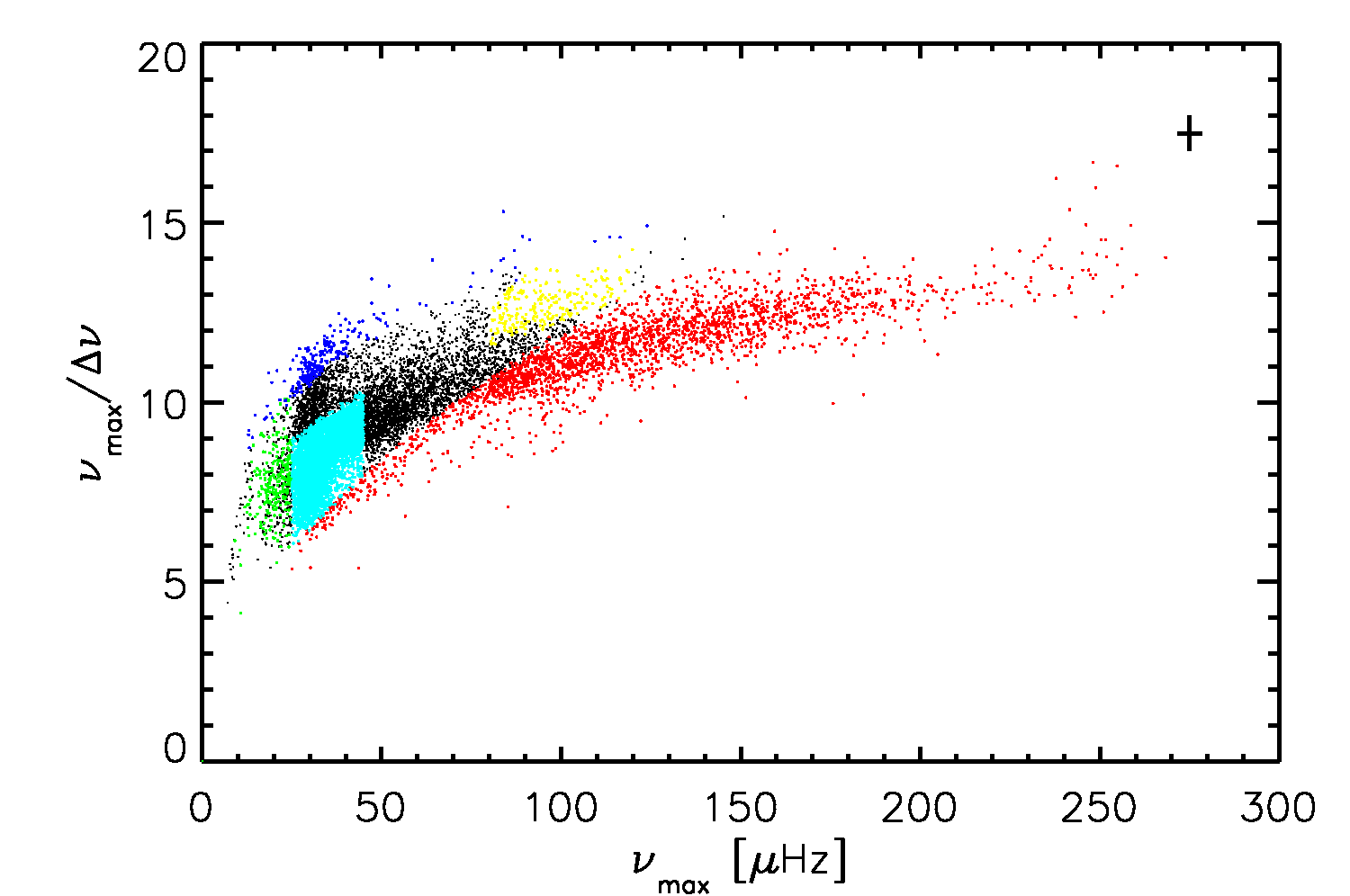}
\end{minipage}
\begin{minipage}{0.45\linewidth}
\centering
\includegraphics[width=\linewidth]{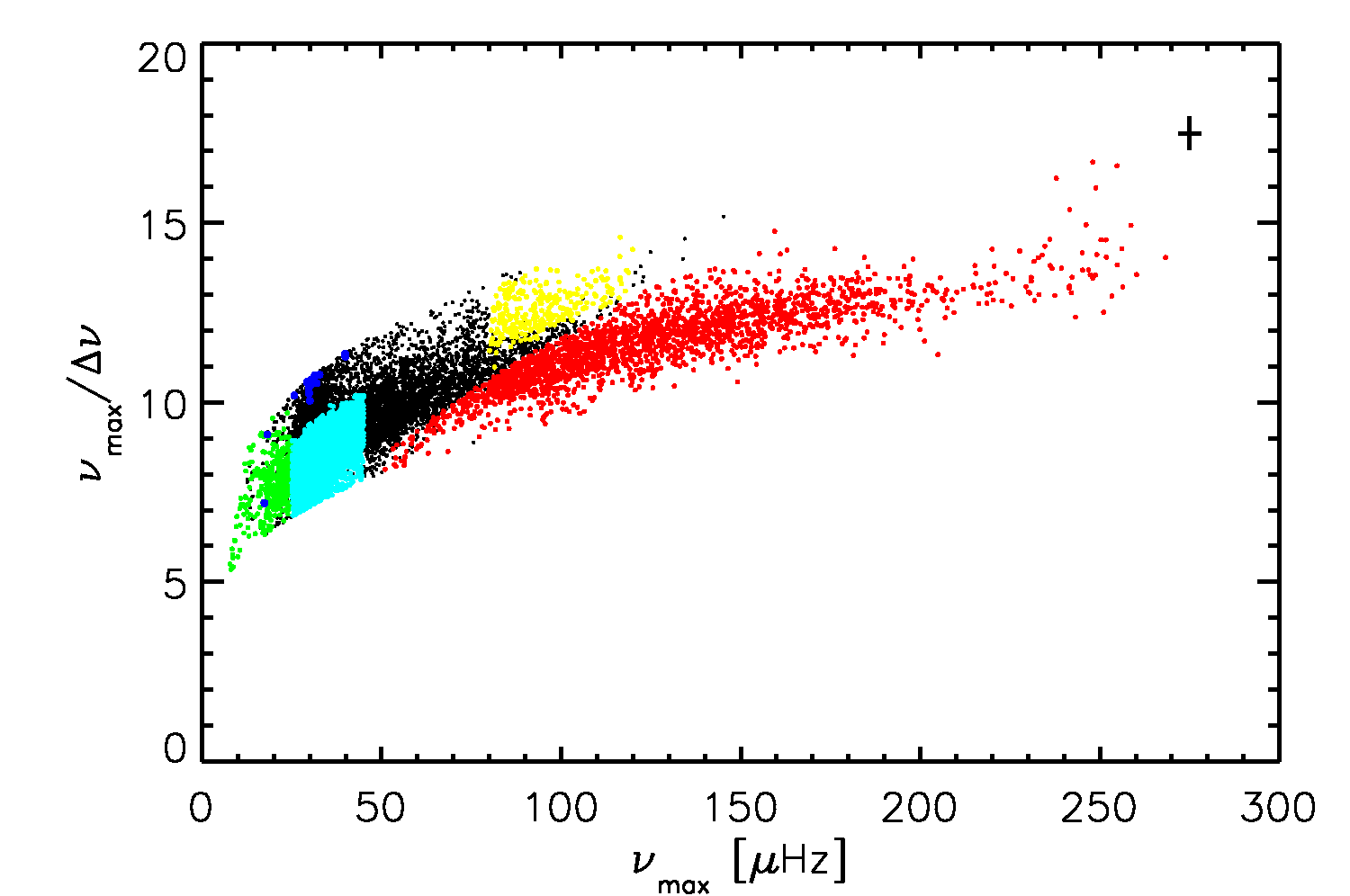}
\end{minipage}
\begin{minipage}{0.45\linewidth}
\centering
\includegraphics[width=\linewidth]{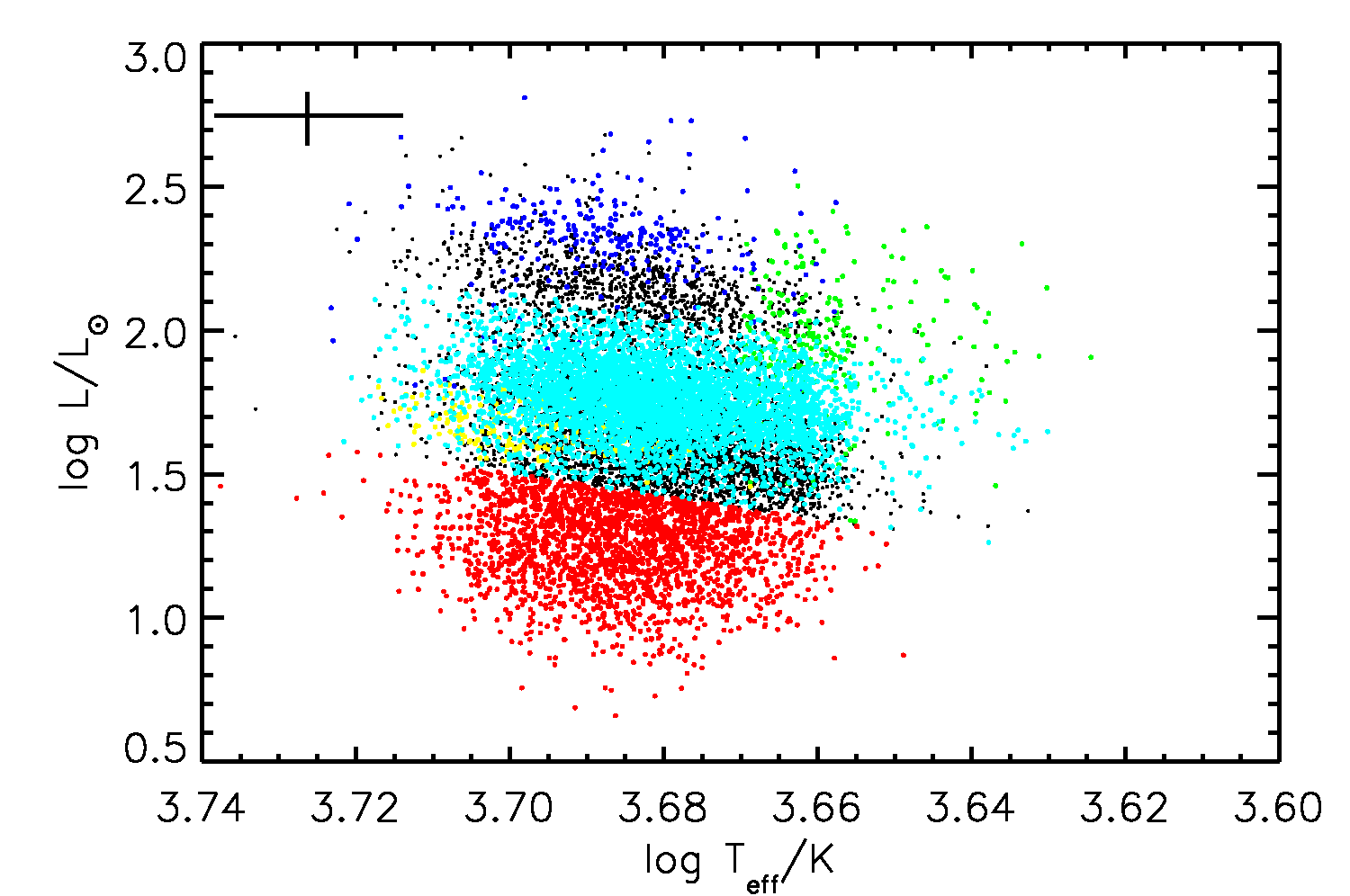}
\end{minipage}
\begin{minipage}{0.45\linewidth}
\centering
\includegraphics[width=\linewidth]{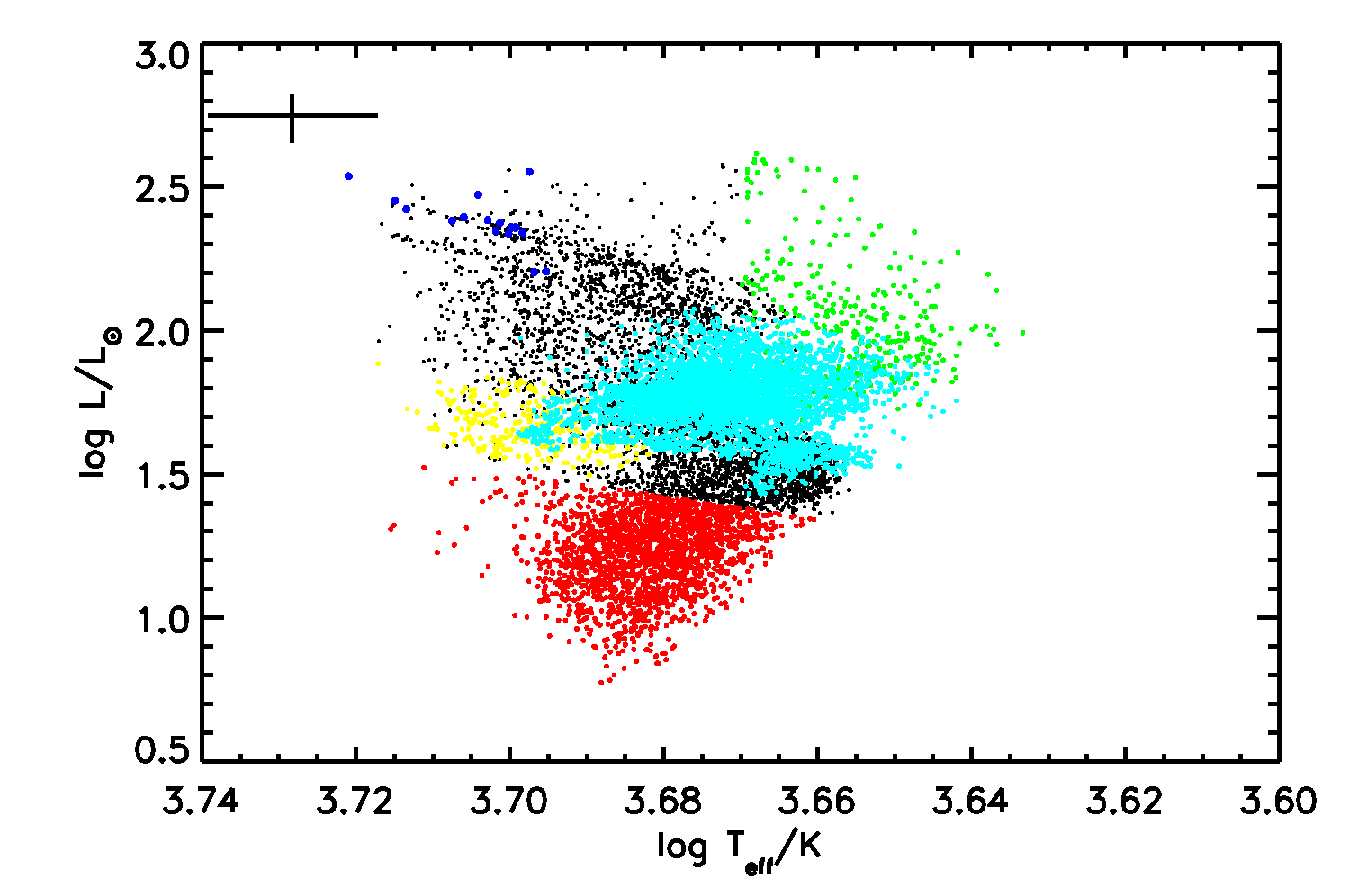}
\end{minipage}
\begin{minipage}{0.45\linewidth}
\centering
\includegraphics[width=\linewidth]{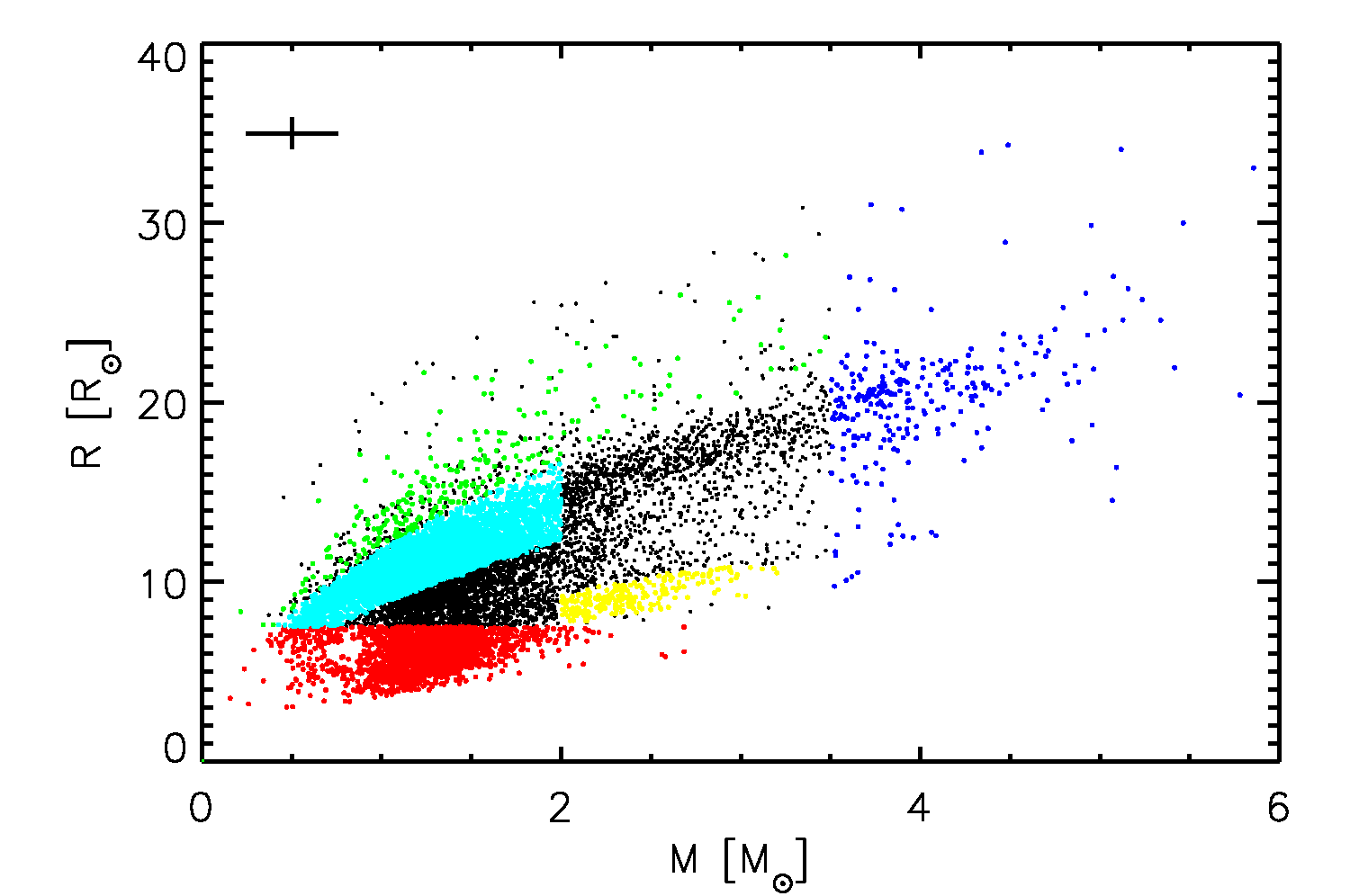}
\end{minipage}
\begin{minipage}{0.45\linewidth}
\centering
\includegraphics[width=\linewidth]{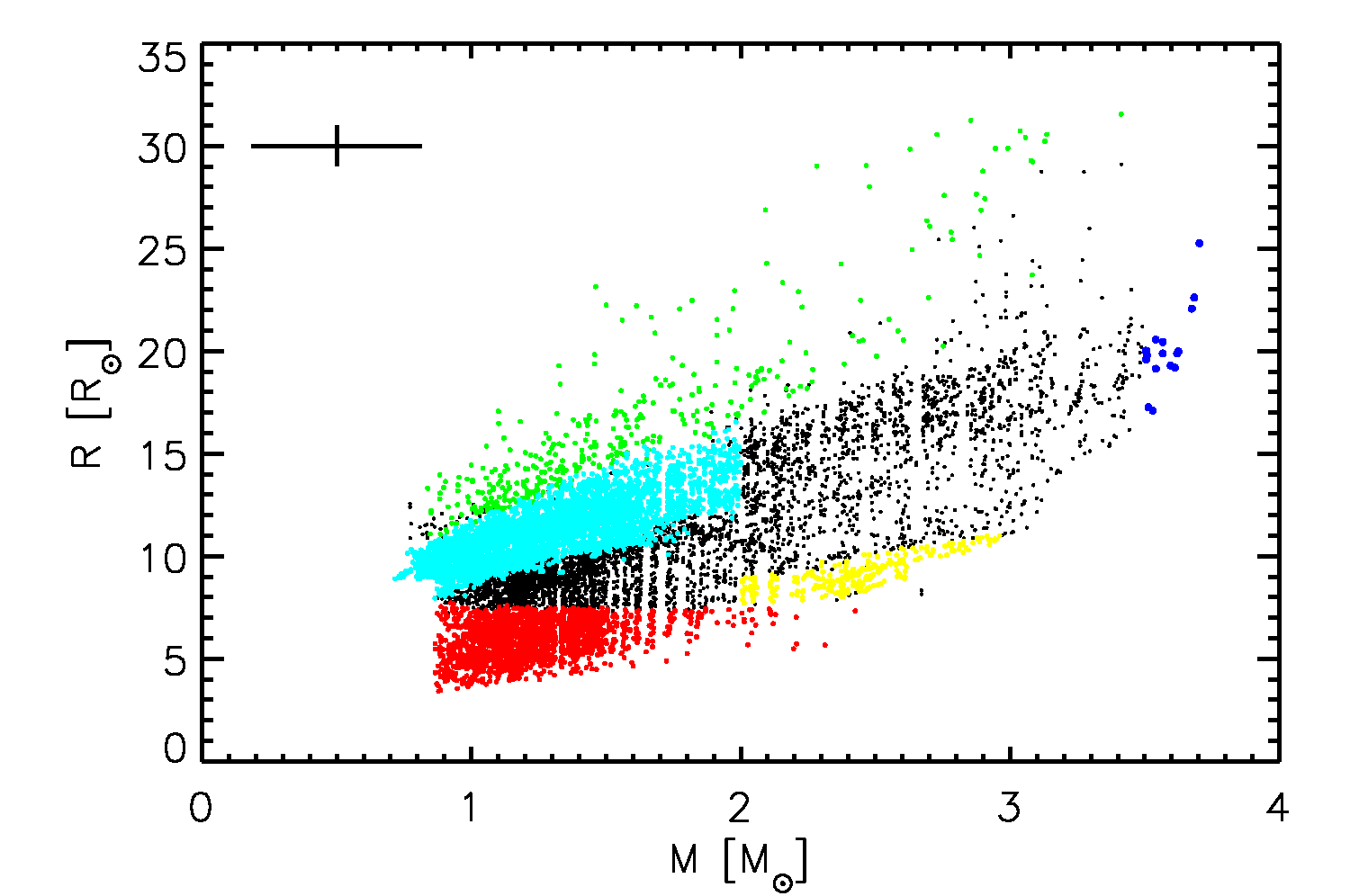}
\end{minipage}
\caption{$\nu_{\rm max}$ versus $\nu_{\rm max}/\meandnu$ diagrams (top), H-R diagrams (centre) and mass versus radius diagrams (bottom) of the public red giants with detected oscillations analysed using the direct method (left) and grid-based approach (right). The different identified populations are indicated with different colours: \textit{red-clump stars} in cyan, \textit{low-luminosity red-giant branch stars} in red \textit{high-luminosity red-giant branch stars} / \textit{asymptotic giant branch stars} in green, \textit{secondary-clump stars} in yellow, and high-mass stars in dark blue. Stars which could not be assigned to a certain population are indicated with black dots. Characteristic uncertainties are indicated with a cross in a corner of each panel. The boundaries of the different populations are approximate.}
\label{popres}
\end{figure*}

We can identify some specific stages of stellar evolution in these diagrams: 
\begin{itemize}
\item The majority of stars with 25~$\mu$Hz~$<$~$\nu_{\rm max}$~$<$~45~$\mu$Hz with masses roughly below 2~M$_{\odot}$ are known to be He-core burning stars that have gone through the Helium flash. These stars have radii of $\sim$10~R$_{\odot}$ and $\log L/\rm L_{\odot}$ of $\sim$2 and are the so-called \textit{red-clump stars} \citep[see e.g.,][for previous detections of the red clump]{miglio2009,huber2010,kallinger2010kepler,mosser2011}. 
\item The stars in the lower radius shoulder in the radius distribution (Fig.~\ref{MRL}) all have low luminosities and lie in the H-R diagram (panel G in Fig.~\ref{res}) below the black dashed line indicating stars with radii of 7.5~R$_{\odot}$. We identify this population as \textit{low-luminosity red-giant branch stars}. 
\item At $\nu_{\rm max}$~$<$~25~$\mu$Hz a group of stars with low effective temperatures is present (see panel B in Fig.~\ref{res}). These stars also have relatively high luminosities (see panel D) and large radii (see panel C), and low to intermediate masses (see panel A). We identify these stars as \textit{high-luminosity red-giant branch stars} or \textit{asymptotic giant branch stars}.
\item There is some evidence for the secondary clump. Stars with masses roughly between 2 and 3~M$_{\odot}$ and radii between 7.5 and 10~R$_{\odot}$ seem to form a separate branch in the mass versus radius diagram with slightly increased temperatures (panel F in Fig.~\ref{res}). These stars are also visible in the $\nu_{\rm max}$ versus $\nu_{\rm max}/\meandnu$ diagram and in the $\nu_{\rm max}$ histogram (Fig.~\ref{dnunumaxhisto}) at the expected location of the secondary clump \citep{girardi1999,huber2010,kallinger2010kepler}. These are stars in their He-burning phase, which are massive enough to have ignited He-burning in a non-degenerate core. We therefore identify these stars as \textit{secondary-clump stars}.
\item The high-mass stars ($\geq$~3.5~M$_{\odot}$) form a distinct group of stars at high $\nu_{\rm max}/\meandnu$ and $\nu_{\rm max}$ values similar to the red-clump stars (see panel A of Fig.~\ref{res}). These stars also have large radii and high luminosities.
\end{itemize}

Additionally, there seems to be a rather steep fall-off for stars with masses greater than roughly 2~M$_{\odot}$ at $\nu_{\rm max}\sim 110$~$\mu$Hz (panel A of Fig.~\ref{res}). This indicates the maximum $\nu_{\rm max}$ of He-burning stars and the region with larger $\nu_{\rm max}$ is populated exclusively with H-shell burning red giant branch stars \citep{miglio2009}. The lack of high-mass stars in the red-giant branch is consistent with high-mass stars evolving much faster than their low-mass counter parts during the H-shell burning phase.

The boundaries of each specific stage of stellar evolution as indicated above are approximate values and could be established thanks to the large sample of stars investigated, despite the relatively large uncertainties in the parameters of individual stars.

The different populations identified above are indicated in the H-R diagram, mass versus radius diagram and $\nu_{\rm max}$ versus $\nu_{\rm max}/\meandnu$ diagram in the left column of Fig.~\ref{popres}. Two distinct unidentified regions are still apparent in the H-R diagram, i.e. the black dots above and below the red clump. The more luminous stars have masses between 2 and 3.5~M$_{\odot}$ and $\nu_{\rm max}$ between 25 and 80~$\mu$Hz, which could be either in the H-shell or He-core/H-shell burning stars. The stars below the red clump are stars with $M$~$<$~2~M$_{\odot}$, $R$~$>$~7.5~R$_{\odot}$ and $\nu_{\rm max}$~$>$~45~$\mu$Hz. These are most likely stars in the H-shell burning phase ascending the red-giant branch.

For the analysis described above we used the KIC effective temperatures and derived masses and radii directly from the scaling relations (Eq.~\ref{numax} and \ref{dnu}). We now compare these results with results from a grid-based approach using information from a grid of models to compute the seismic effective temperatures, masses, radii and luminosities of the stars \citep{kallinger2010kepler}. In the grid-based approach stellar evolution is taken into account, while the scaling relations assume that all values of $T_{\rm eff}$ are possible for a star of a given mass and radius. Using the grid-based approach should therefore reduce the uncertainties of the determined parameters \citep[see also][]{gai2010}. The results from the grid-based approach are shown in the right column of Fig.~\ref{popres}. The grid-based approach indeed improves the structure in the H-R diagram, mass versus radius diagram and $\nu_{\rm max}$ versus $\nu_{\rm max}/\meandnu$ diagram, but does not alter the identification of the different populations, except for the high-mass population. The stars with masses between 4 and 6~M$_{\odot}$ as computed from the direct method have all been corrected to have masses with $M$~$<$~4~M$_{\odot}$ and the high-mass stars population has been reduced. These corrections indicate that for the stars with highest masses the uncertainties in the KIC and the direct method overestimate the stellar masses.

\subsection{Effectiveness of oscillation detection}
\begin{figure}
\begin{minipage}{\linewidth}
\centering
\includegraphics[width=\linewidth]{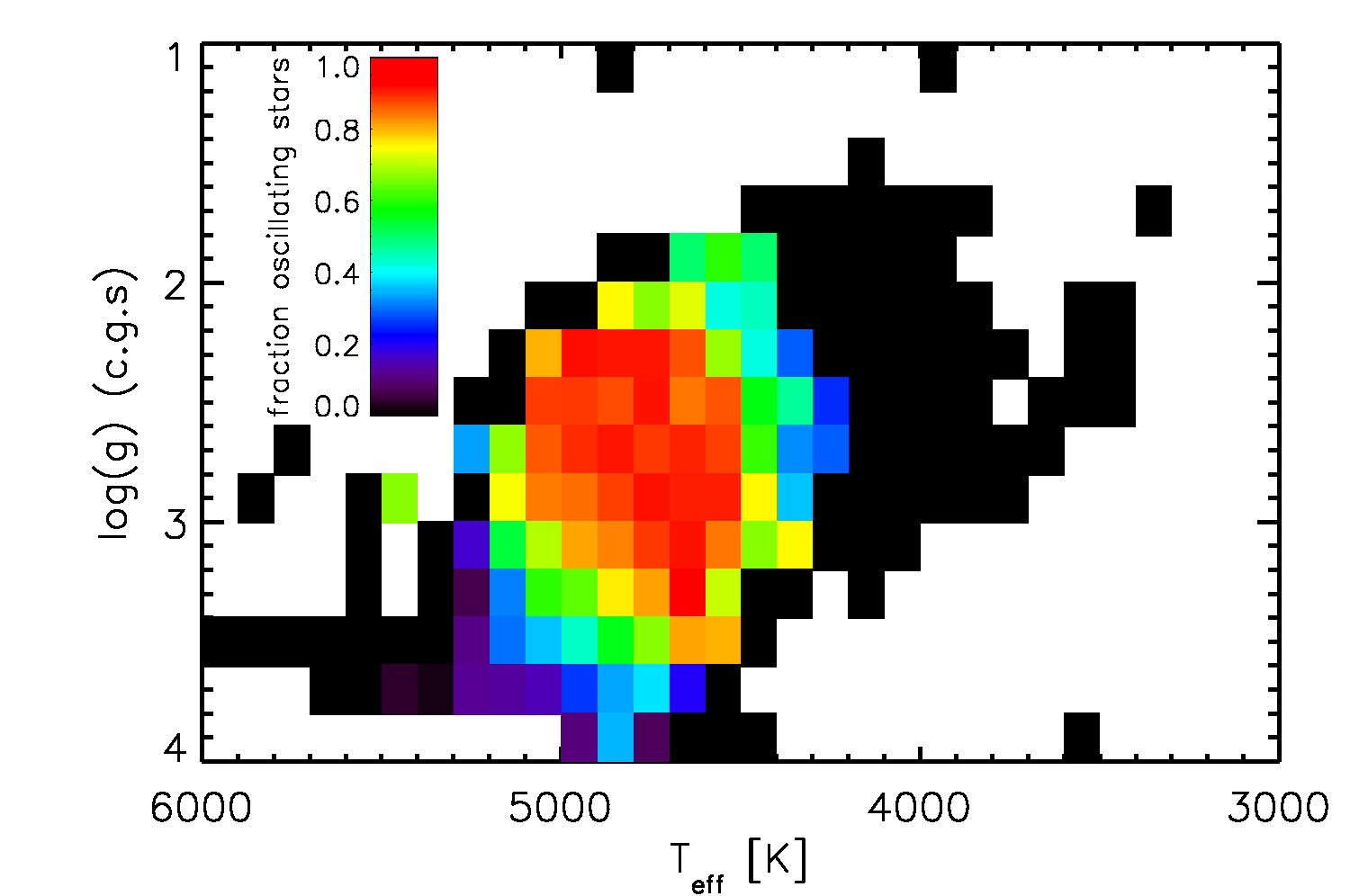}
\end{minipage}
\caption{Zoom of Fig.~\ref{loggteff} into the red-giant range showing the fraction of red giants with detected solar-like oscillations in $\log g$ versus $T_{\rm eff}$. Each square represents an interval of 0.2~dex in $\log g$ and 100~K in $T_{\rm eff}$.} 
\label{fracloggteff}
\end{figure}

The fraction of red giants for which oscillations have been detected has its maximum at roughly $\log g$~=~2.7 and $T_{\rm eff}$~=~4800~K and decreases for higher and lower values of $\log g$ and $T_{\rm eff}$ (see Fig.~\ref{fracloggteff}). There is a notable absence of stars with detected oscillations in the region with $T_{\rm eff}$~$<$~4200~K (right top corner in Fig.~\ref{fracloggteff}). These are the coolest stars with lowest $\log g$. These stars have very large radii and are expected to have long oscillation periods, for which (as stated above) the timespan of the Q1 data are not sufficient. It is also clear that for hotter stars ($T_{\rm eff}$~$>$~5200~K) with higher gravity ($\log g$~$>$~3) less oscillating stars have been detected. These stars are in a less evolved state, in which possible solar-like oscillations occur at frequencies higher than the Nyquist frequency (283~$\mu$Hz) of the long cadence \textit{Kepler} data, investigated here. 
We also checked the fraction of detections as a function of magnitude. This shows no clear trend. The ratios of stars in each magnitude interval with detected oscillations to stars without detections are very similar. This shows that the detection of solar-like oscillations in red-giant stars does not depend (strongly) on the apparent magnitude. Note that there are only a few red giants brighter than 8$^{\rm th}$ magnitude and no red-giant stars fainter than 14$^{\rm th}$ magnitude present among the public data. 

\begin{figure}
\begin{minipage}{\linewidth}
\centering
\includegraphics[width=\linewidth]{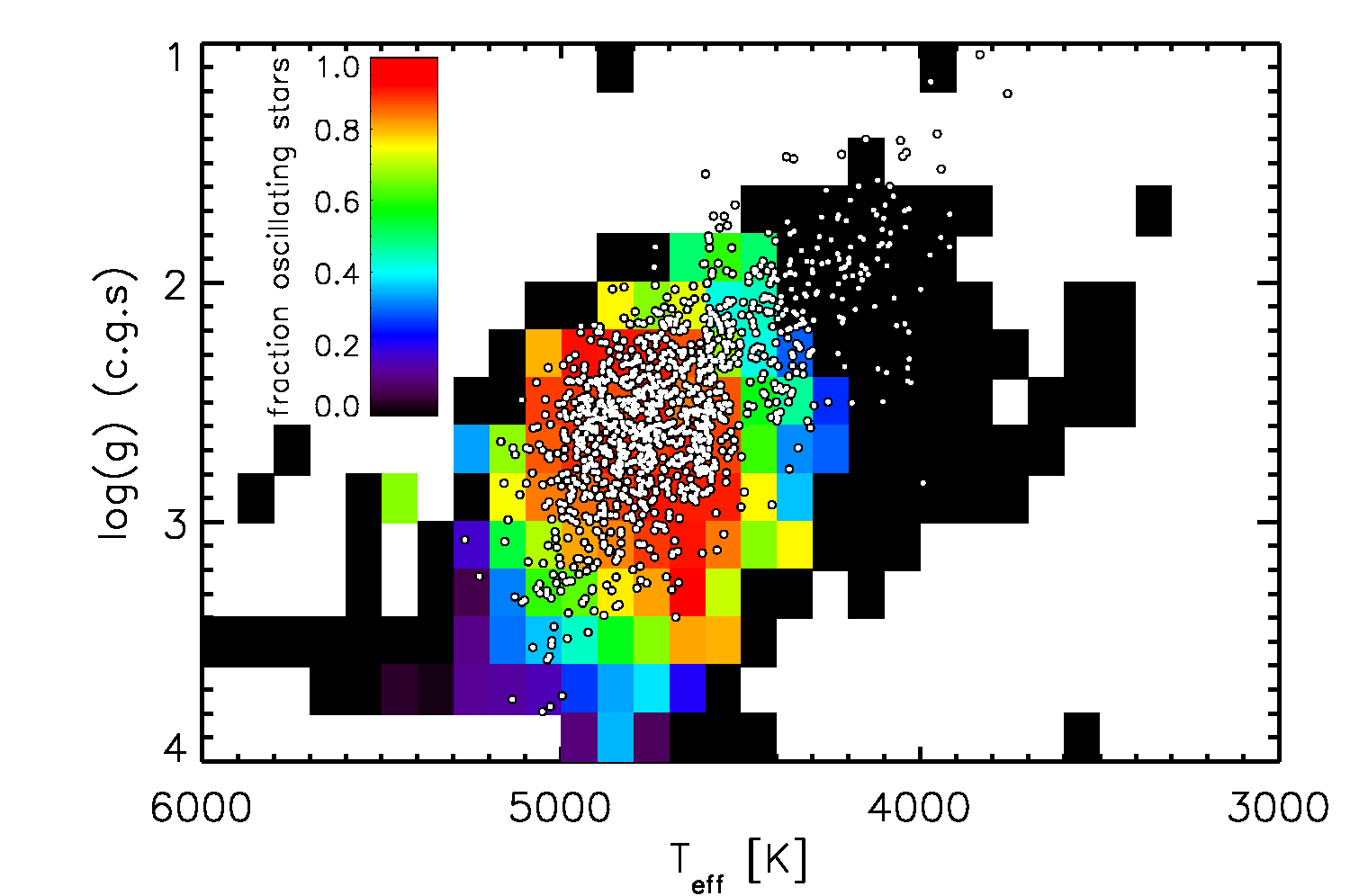}
\end{minipage}
\caption{Same as Fig.~\ref{fracloggteff} with the location of the red-giant stars analysed within the Kepler Asteroseismic Science Consortium (see text) shown (white dots).} 
\label{fracloggteffkasc}
\end{figure}

We have compared the population of red giants in the public data described here with the red giants analysed within the Kepler Asteroseismic Science Consortium (KASC) \citep{bedding2010,hekker2010comp,huber2010,kallinger2010kepler}. The KASC stars are shown with white dots in Fig.~\ref{fracloggteffkasc}. It is clear that a large fraction of these stars are located in the same $T_{\rm eff}$ - $\log g$ range as the public data of stars showing oscillations, although the lower right corner of the red/orange region indicating a high fraction ($>$80\%) of detected oscillators in the public data is not well populated with KASC stars. There are also many fewer oscillating KASC red giants with $\log g>$~3.5. The lack of KASC stars in both regions could possibly be caused by selection effects, and low number statistics due to the fact that in these regions in the H-R diagram stars evolve relatively rapidly and thus that the chance of observing a star in this region is lower. Indeed the astrometric control set, the largest component of KASC  
analysed red giants, was selected to have large distances, hence low $\log g$. Finally, in the low $\log g$ - low $T_{\rm eff}$ range where hardly any oscillations have been detected in the public data, we see a non-negligible number of oscillating KASC stars. This is most likely due to the fact that the KASC timeseries analysed are three months longer than the public data analysed here. 

Comparing our results with the results of the two CoRoT fields \citep{mosser2010} shows that the distributions of $\nu_{\rm max}$ and $\meandnu$ peak at very similar values, although the secondary clump is much more pronounced in the large public \textit{Kepler} sample investigated here (see also their Fig. 16). We also see that a more significant fraction of the stars in this work have masses above two M$_{\odot}$, while the distribution of the radii is similar for the CoRoT and \textit{Kepler} samples.



\subsection{Surface gravity}
\begin{figure}
\begin{minipage}{\linewidth}
\centering
\includegraphics[width=\linewidth]{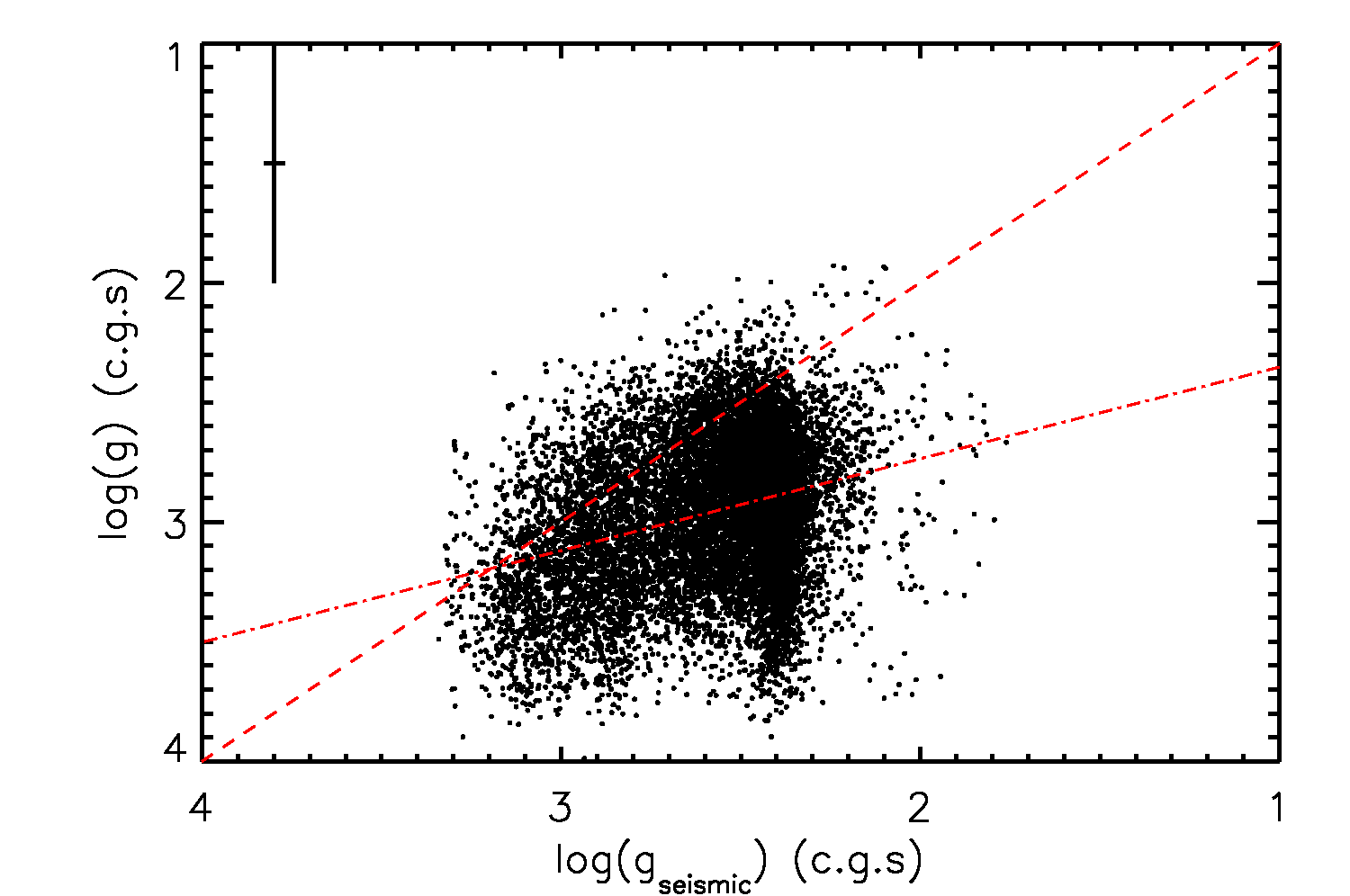}
\end{minipage}
\caption{$\log g$ of the \textit{Kepler} Input Catalogue as a function of $\log g_{\rm seismic}$ determined from asteroseismic parameters. Typical uncertainties are indicated in the left top corner. The red dashed line shows the one-to-one relation and the red dashed-dotted line a linear fit through the data. Note that the x- and y-axis have the same range to illustrate the narrow distribution of $\log g_{\rm seismic}$ compared to $\log g$ of the \textit{Kepler} Input Catalogue.}
\label{logg}
\end{figure}

For the analyses carried out in this work, $\log g$ values from the \textit{Kepler} Input Catalogue have been used, because those values are available for stars with and without detected oscillations. It is however interesting to compare the KIC $\log g$ with $\log g$ computed from asteroseismology. Figure~\ref{logg} shows the KIC $\log g $ as a function of $\log g_{\rm seismic} $ with $g_{\rm seismic} \propto T_{\rm eff}^{0.5}\nu_{\rm max}$, in which $T_{\rm eff}$ values from the KIC have been used.
This Figure shows that the values in the KIC are in general consistent or higher than the values obtained using asteroseismology. The spread in $\log g_{\rm seismic}$ is notably smaller than in the KIC values with the majority of the stars having $\log g_{\rm seismic}$ roughly between 2.3 and 2.7. This concentration indicates the red clump with He-core burning stars, while the stars with higher $\log g_{\rm seismic}$, i.e., $>$~2.7 are the \textit{low-luminosity red-giant branch stars} and the stars with lower $\log g_{\rm seismic}$, i.e., $<$~2.2, the \textit{high-luminosity red-giant branch stars} or \textit{asymptotic giant branch stars}.

Note that the uncertainties in the results for $\log g_{\rm seismic} $ are considerably smaller than the uncertainties in the KIC $\log g$ values. This has been predicted by \citet{gai2010}. They show that $\log g_{\rm seismic}$ is a very robust parameter with uncertainties of only a few percent.


\section{Summary}
The 33~days of public \textit{Kepler} data observed in Q1 have been investigated for solar-like oscillations in red-giant stars. For 71\% of the red giants, oscillations could be detected and for these stars asteroseismic masses, radii, luminosities and distances have been derived. These are computed directly from scaling relations (Eqs.~\ref{numax}~and~\ref{dnu}) and in a second analysis using a grid-based approach \citep{stello2009R,kallinger2010kepler,gai2010,basu2010}. 

The stellar parameters determined with the direct method from the relatively short timeseries are accurate enough to distinguish between different populations of stars, i.e., \textit{low-luminosity red-giant branch stars} and \textit{high-luminosity red-giant branch stars} in the H-shell burning phase or \textit{asymptotic giant branch stars}, He-core burning \textit{clump} and \textit{secondary-clump} stars. Furthermore, the lack of high-mass stars with high $\nu_{\rm max}$ ($>$~110~$\mu$Hz) confirms observationally the theoretically known difference in evolutionary timescales between stars of different masses.

The fact that we did not detect oscillations for 29\% of the red giants in the sample can be explained by observational influences, such as the limited timespan of the data and instrumental artefacts. No correlation with detectability could be determined with the apparent magnitude of the stars. Over the next few years when more data will become available and minor problems with instrumental artefacts overcome, we expect to observe solar-like oscillations in a larger fraction of the giants. CoRoT observations spanning about 150~days have shown that more than 75\% of the red-giant candidates with apparent magnitude brighter than 13$^{\rm th}$ magnitude show oscillations \citep{mosser2011}. 


\section*{acknowledgements}
The authors gratefully acknowledge the \textit{Kepler} Science Team and all those who have contributed to making the \textit{Kepler} mission possible. Funding for this Discovery mission is provided by NASA's Science Mission Directorate. SH, YE and WJC acknowledge financial support from the UK Science and Technology Facilities Council (STFC). SH also acknowledges financial support from the Netherlands Organisation for Scientific Research (NWO). We would like to thank the anonymous referee for useful comments which improved the manuscript considerably.


\begin{thebibliography}{}

\bibitem[\protect\citeauthoryear{{Basu}, {Grundahl}, {Stello}, {Kallinger},
  {Hekker}, {Mosser}, {Garc{\'{\i}}a}, {Mathur}, {Brogaard}, {Bruntt},
  {Chaplin} \& {et al.}}{{Basu} et~al.}{2011}]{basu2010}
{Basu} S.,  {Grundahl} F.,  {Stello} D.,  {Kallinger} T.,  {Hekker} S.,
  {Mosser} B.,  {Garc{\'{\i}}a} R.~A.,  {Mathur} S.,  {Brogaard} K.,  {Bruntt}
  H.,  {Chaplin} W.~J.,    {et al.} 2011, ApJ, 729, L10

\bibitem[\protect\citeauthoryear{{Bedding}, {Huber}, {Stello}, {Elsworth},
  {Hekker}, {Kallinger}, {Mathur}, {Mosser}, {Preston}, {Ballot}, {Barban},
  {Broomhall} \& {et al.}}{{Bedding} et~al.}{2010}]{bedding2010}
{Bedding} T.~R.,  {Huber} D.,  {Stello} D.,  {Elsworth} Y.~P.,  {Hekker} S.,
  {Kallinger} T.,  {Mathur} S.,  {Mosser} B.,  {Preston} H.~L.,  {Ballot} J.,
  {Barban} C.,  {Broomhall} A.~M.,    {et al.} 2010, ApJ, 713, L176

\bibitem[\protect\citeauthoryear{{Borucki}, {Koch}, {Batalha}, {Caldwell},
  {Christensen-Dalsgaard}, {Cochran}, {Dunham}, {Gautier}, {Geary},
  {Gilliland}, {Jenkins}, {Kjeldsen}, {Lissauer} \& {Rowe}}{{Borucki}
  et~al.}{2009}]{borucki2009}
{Borucki} W.,  {Koch} D.,  {Batalha} N.,  {Caldwell} D.,
  {Christensen-Dalsgaard} J.,  {Cochran} W.~D.,  {Dunham} E.,  {Gautier} T.~N.,
   {Geary} J.,  {Gilliland} R.,  {Jenkins} J.,  {Kjeldsen} H.,  {Lissauer}
  J.~J.,    {Rowe} J.,  2009, in IAU Symposium Vol.~253 of IAU Symposium,
  {KEPLER: Search for Earth-Size Planets in the Habitable Zone}.
pp 289--299

\bibitem[\protect\citeauthoryear{{Brown}, {Latham}, {Everett} \&
  {Esquerdo}}{{Brown} et~al.}{2011}]{brown2011}
{Brown} T.~M.,  {Latham} D.~W.,  {Everett} M.~E.,    {Esquerdo} G.~A.,  2011,
  ArXiv e-prints:1102.0342

\bibitem[\protect\citeauthoryear{{Bruntt}, {Frandsen} \& {Thygesen}}{{Bruntt}
  et~al.}{2010}]{bruntt2011}
{Bruntt} H.,  {Frandsen} S.,    {Thygesen} A.~O.,  2010, ArXiv e-prints: 1012.0436

\bibitem[\protect\citeauthoryear{{Carrier}, {De Ridder}, {Baudin}, {Barban},
  {Hatzes}, {Hekker}, {Kallinger}, {Miglio}, {Montalb{\'a}n}, {Morel}, {Weiss},
  {Auvergne}, {Baglin}, {Catala}, {Michel} \& {Samadi}}{{Carrier}
  et~al.}{2010}]{carrier2010}
{Carrier} F.,  {De Ridder} J.,  {Baudin} F.,  {Barban} C.,  {Hatzes} A.~P.,
  {Hekker} S.,  {Kallinger} T.,  {Miglio} A.,  {Montalb{\'a}n} J.,  {Morel} T.,
   {Weiss} W.~W.,  {Auvergne} M.,  {Baglin} A.,  {Catala} C.,  {Michel} E.,
  {Samadi} R.,  2010, A\&A, 509, A73

\bibitem[\protect\citeauthoryear{{Ciardi}, {von Braun}, {Bryden}, {van Eyken},
  {Howell}, {Kane}, {Plavchan} \& {Stauffer}}{{Ciardi}
  et~al.}{2010}]{ciardi2010}
{Ciardi} D.~R.,  {von Braun} K.,  {Bryden} G.,  {van Eyken} J.,  {Howell}
  S.~B.,  {Kane} S.~R.,  {Plavchan} P.,    {Stauffer} J.~R.,  2010, ArXiv
  e-prints:1009.1840

\bibitem[\protect\citeauthoryear{{De Ridder}, {Barban}, {Baudin}, {Carrier},
  {Hatzes}, {Hekker}, {Kallinger}, {Weiss}, {Baglin}, {Auvergne}, {Samadi},
  {Barge} \& {Deleuil}}{{De Ridder} et~al.}{2009}]{deridder2009}
{De Ridder} J.,  {Barban} C.,  {Baudin} F.,  {Carrier} F.,  {Hatzes} A.~P.,
  {Hekker} S.,  {Kallinger} T.,  {Weiss} W.~W.,  {Baglin} A.,  {Auvergne} M.,
  {Samadi} R.,  {Barge} P.,    {Deleuil} M.,  2009, Nature, 459, 398

\bibitem[\protect\citeauthoryear{{Gai}, {Basu}, {Chaplin} \& {Elsworth}}{{Gai}
  et~al.}{2010}]{gai2010}
{Gai} N.,  {Basu} S.,  {Chaplin} W.~J.,    {Elsworth} Y.,  2010, ArXiv
  e-prints:1009.3018

\bibitem[\protect\citeauthoryear{{Girardi}}{{Girardi}}{1999}]{girardi1999}
{Girardi} L.,  1999, MNRAS, 308, 818

\bibitem[\protect\citeauthoryear{{Hekker}}{{Hekker}}{2010}]{hekker2010AN}
{Hekker} S.,  2010, Astronomische Nachrichten, 331, 1004

\bibitem[\protect\citeauthoryear{{Hekker}, {Barban}, {Baudin}, {De Ridder},
  {Kallinger}, {Morel}, {Chaplin} \& {Elsworth}}{{Hekker}
  et~al.}{2010a}]{hekker2010pb}
{Hekker} S.,  {Barban} C.,  {Baudin} F.,  {De Ridder} J.,  {Kallinger} T.,
  {Morel} T.,  {Chaplin} W.~J.,    {Elsworth} Y.,  2010a, A\&A, 520, A60

\bibitem[\protect\citeauthoryear{{Hekker}, {Basu}, {Stello}, {Kallinger} \&
  {Grundahl}}{{Hekker} et~al.}{2011a}]{hekker2010wg2}
{Hekker} S.,  {Basu} S.,  {Stello} D.,  {Kallinger} T.,    {Grundahl} F.,
  2011a, A\&A, submitted

\bibitem[\protect\citeauthoryear{{Hekker}, {Broomhall}, {Chaplin}, {Elsworth},
  {Fletcher}, {New}, {Arentoft}, {Quirion} \& {Kjeldsen}}{{Hekker}
  et~al.}{2010b}]{hekker2010pipe}
{Hekker} S.,  {Broomhall} A.,  {Chaplin} W.~J.,  {Elsworth} Y.~P.,  {Fletcher}
  S.~T.,  {New} R.,  {Arentoft} T.,  {Quirion} P.,    {Kjeldsen} H.,  2010b,
  MNRAS, 402, 2049

\bibitem[\protect\citeauthoryear{{Hekker}, {Debosscher}, {Huber}, {Hidas}, {De
  Ridder}, {Aerts}, {Stello}, {Bedding}, {Gilliland}, {Christensen-Dalsgaard},
  {Brown}, {Kjeldsen}, {Borucki}, {Koch} \& {Jenkins}}{{Hekker}
  et~al.}{2010c}]{hekker2010bin}
{Hekker} S.,  {Debosscher} J.,  {Huber} D.,  {Hidas} M.~G.,  {De Ridder} J.,
  {Aerts} C.,  {Stello} D.,  {Bedding} T.~R.,  {Gilliland} R.~L.,
  {Christensen-Dalsgaard} J.,  {Brown} T.~M.,  {Kjeldsen} H.,  {Borucki} W.~J.,
   {Koch} D.,    {Jenkins} J.~M.,  2010c, ApJ, 713, L187

\bibitem[\protect\citeauthoryear{{Hekker}, {Elsworth}, {De Ridder}, {Mosser},
  {Garc{\'{\i}}a}, {Kallinger}, {Mathur}, {Huber}, {Buzasi}, {Preston}, {Hale},
  {Ballot}, {Chaplin}, {R{\'e}gulo}, {Bedding}, {Stello} \& {et al.}}{{Hekker}
  et~al.}{2011b}]{hekker2010comp}
{Hekker} S.,  {Elsworth} Y.,  {De Ridder} J.,  {Mosser} B.,  {Garc{\'{\i}}a}
  R.~A.,  {Kallinger} T.,  {Mathur} S.,  {Huber} D.,  {Buzasi} D.~L.,
  {Preston} H.~L.,  {Hale} S.~J.,  {Ballot} J.,  {Chaplin} W.~J.,  {R{\'e}gulo}
  C.,  {Bedding} T.~R.,  {Stello} D.,    {et al.} 2011b, A\&A, 525, A131

\bibitem[\protect\citeauthoryear{{Hekker}, {Kallinger}, {Baudin}, {De Ridder},
  {Barban}, {Carrier}, {Hatzes}, {Weiss} \& {Baglin}}{{Hekker}
  et~al.}{2009}]{hekker2009}
{Hekker} S.,  {Kallinger} T.,  {Baudin} F.,  {De Ridder} J.,  {Barban} C.,
  {Carrier} F.,  {Hatzes} A.~P.,  {Weiss} W.~W.,    {Baglin} A.,  2009, A\&A,
  506, 465

\bibitem[\protect\citeauthoryear{{Huber}, {Bedding}, {Stello}, {Mosser},
  {Mathur}, {Kallinger}, {Hekker}, {Elsworth}, {Buzasi}, {De Ridder},
  {Gilliland} \& {et al.}}{{Huber} et~al.}{2010}]{huber2010}
{Huber} D.,  {Bedding} T.~R.,  {Stello} D.,  {Mosser} B.,  {Mathur} S.,
  {Kallinger} T.,  {Hekker} S.,  {Elsworth} Y.~P.,  {Buzasi} D.~L.,  {De
  Ridder} J.,  {Gilliland} R.~L.,    {et al.} 2010, ApJ, 723, 1607

\bibitem[\protect\citeauthoryear{{Jenkins}, {Caldwell}, {Chandrasekaran},
  {Twicken}, {Bryson}, {Quintana}, {Clarke}, {Li}, {Allen}, {Tenenbaum}, {Wu},
  {Klaus}, {Middour} \& {et al.}}{{Jenkins} et~al.}{2010}]{jenkins2010a}
{Jenkins} J.~M.,  {Caldwell} D.~A.,  {Chandrasekaran} H.,  {Twicken} J.~D.,
  {Bryson} S.~T.,  {Quintana} E.~V.,  {Clarke} B.~D.,  {Li} J.,  {Allen} C.,
  {Tenenbaum} P.,  {Wu} H.,  {Klaus} T.~C.,  {Middour} C.~K.,    {et al.} 2010,
  ApJ, 713, L87

\bibitem[\protect\citeauthoryear{{Kallinger}, {Mosser}, {Hekker}, {Huber},
  {Stello}, {Mathur}, {Basu}, {Bedding}, {Chaplin}, {De Ridder}, {Elsworth} \&
  {et al.}}{{Kallinger} et~al.}{2010a}]{kallinger2010kepler}
{Kallinger} T.,  {Mosser} B.,  {Hekker} S.,  {Huber} D.,  {Stello} D.,
  {Mathur} S.,  {Basu} S.,  {Bedding} T.~R.,  {Chaplin} W.~J.,  {De Ridder} J.,
   {Elsworth} Y.~P.,    {et al.} 2010a, A\&A, 522, A1

\bibitem[\protect\citeauthoryear{{Kallinger}, {Weiss}, {Barban}, {Baudin},
  {Cameron}, {Carrier}, {De Ridder}, {Goupil}, {Gruberbauer}, {Hatzes},
  {Hekker}, {Samadi} \& {Deleuil}}{{Kallinger}
  et~al.}{2010b}]{kallinger2010corot}
{Kallinger} T.,  {Weiss} W.~W.,  {Barban} C.,  {Baudin} F.,  {Cameron} C.,
  {Carrier} F.,  {De Ridder} J.,  {Goupil} M.,  {Gruberbauer} M.,  {Hatzes} A.,
   {Hekker} S.,  {Samadi} R.,    {Deleuil} M.,  2010b, A\&A, 509, A77

\bibitem[\protect\citeauthoryear{{Kjeldsen} \& {Bedding}}{{Kjeldsen} \&
  {Bedding}}{1995}]{kjeldsen1995}
{Kjeldsen} H.,  {Bedding} T.~R.,  1995, A\&A, 293, 87

\bibitem[\protect\citeauthoryear{{Miglio}, {Montalb{\'a}n}, {Baudin},
  {Eggenberger}, {Noels}, {Hekker}, {De Ridder}, {Weiss} \& {Baglin}}{{Miglio}
  et~al.}{2009}]{miglio2009}
{Miglio} A.,  {Montalb{\'a}n} J.,  {Baudin} F.,  {Eggenberger} P.,  {Noels} A.,
   {Hekker} S.,  {De Ridder} J.,  {Weiss} W.,    {Baglin} A.,  2009, A\&A, 503,
  L21

\bibitem[\protect\citeauthoryear{{Miglio}, {Montalb{\'a}n}, {Carrier}, {De
  Ridder}, {Mosser}, {Eggenberger}, {Scuflaire}, {Ventura}, {D'Antona}, {Noels}
  \& {Baglin}}{{Miglio} et~al.}{2010}]{miglio2010}
{Miglio} A.,  {Montalb{\'a}n} J.,  {Carrier} F.,  {De Ridder} J.,  {Mosser} B.,
   {Eggenberger} P.,  {Scuflaire} R.,  {Ventura} P.,  {D'Antona} F.,  {Noels}
  A.,    {Baglin} A.,  2010, A\&A, 520, L6

\bibitem[\protect\citeauthoryear{{Mosser}, {Belkacem}, {Goupil}, {Miglio},
  {Morel}, {Barban}, {Baudin}, {Hekker}, {Samadi}, {De Ridder}, {Weiss},
  {Auvergne} \& {Baglin}}{{Mosser} et~al.}{2010}]{mosser2010}
{Mosser} B.,  {Belkacem} K.,  {Goupil} M.,  {Miglio} A.,  {Morel} T.,  {Barban}
  C.,  {Baudin} F.,  {Hekker} S.,  {Samadi} R.,  {De Ridder} J.,  {Weiss} W.,
  {Auvergne} M.,    {Baglin} A.,  2010, A\&A, 517, A22

\bibitem[\protect\citeauthoryear{{Mosser}, {Belkacem}, {Goupil}, {Michel},
  {Elsworth}, {Barban}, {Kallinger}, {Hekker}, {De Ridder}, {Samadi}, {Baudin},
  {Pinheiro}, {Auvergne}, {Baglin} \& {Catala}}{{Mosser}
  et~al.}{2011}]{mosser2011}
{Mosser} B.,  {Belkacem} K.,  {Goupil} M.~J.,  {Michel} E.,  {Elsworth} Y.,
  {Barban} C.,  {Kallinger} T.,  {Hekker} S.,  {De Ridder} J.,  {Samadi} R.,
  {Baudin} F.,  {Pinheiro} F.~J.~G.,  {Auvergne} M.,  {Baglin} A.,    {Catala}
  C.,  2011, A\&A, 525, L9

\bibitem[\protect\citeauthoryear{{Pr{\v s}a}, {Batalha}, {Slawson}, {Doyle},
  {Welsh}, {Orosz}, {Seager}, {Rucker}, {Mjaseth}, {Engle}, {Conroy},
  {Jenkins}, {Caldwell}, {Koch} \& {Borucki}}{{Pr{\v s}a}
  et~al.}{2011}]{prsa2010}
{Pr{\v s}a} A.,  {Batalha} N.,  {Slawson} R.~W.,  {Doyle} L.~R.,  {Welsh}
  W.~F.,  {Orosz} J.~A.,  {Seager} S.,  {Rucker} M.,  {Mjaseth} K.,  {Engle}
  S.~G.,  {Conroy} K.,  {Jenkins} J.,  {Caldwell} D.,  {Koch} D.,    {Borucki}
  W.,  2011, AJ, 141, 83

\bibitem[\protect\citeauthoryear{{Stello}, {Chaplin}, {Basu}, {Elsworth} \&
  {Bedding}}{{Stello} et~al.}{2009a}]{stello2009}
{Stello} D.,  {Chaplin} W.~J.,  {Basu} S.,  {Elsworth} Y.,    {Bedding} T.~R.,
  2009a, MNRAS, 400, L80

\bibitem[\protect\citeauthoryear{{Stello}, {Chaplin}, {Bruntt}, {Creevey},
  {Garc{\'{\i}}a-Hern{\'a}ndez}, {Monteiro}, {Moya}, {Quirion}, {Sousa},
  {Su{\'a}rez} \& {et al.}}{{Stello} et~al.}{2009b}]{stello2009R}
{Stello} D.,  {Chaplin} W.~J.,  {Bruntt} H.,  {Creevey} O.~L.,
  {Garc{\'{\i}}a-Hern{\'a}ndez} A.,  {Monteiro} M.~J.~P.~F.~G.,  {Moya} A.,
  {Quirion} P.,  {Sousa} S.~G.,  {Su{\'a}rez} J.,    {et al.} 2009b, ApJ, 700,
  1589

\bibitem[\protect\citeauthoryear{{Verner}, {Elsworth}, {Chaplin}, {Campante},
  {Corsaro}, {Gaulme}, {Hekker}, {Huber}, {Karoff}, {Mathur}, {Mosser} \& {et
  al}}{{Verner} et~al.}{2011}]{verner2011}
{Verner} G.,  {Elsworth} Y.,  {Chaplin} W.~J.,  {Campante} T.,  {Corsaro} E.,
  {Gaulme} P.,  {Hekker} S.,  {Huber} D.,  {Karoff} C.,  {Mathur} S.,  {Mosser}
  B.,    {et al} 2011, MNRAS, submitted

\end{thebibliography}


\label{lastpage}

\end{document}